%% file: main.tex
\documentclass[twocolumn,twocolappendix,resetfootnote]{aastex701}

\usepackage{hyperref}
\usepackage{amsmath}
\shorttitle{Search for $z\sim10$ LRDs}
\shortauthors{T. Tanaka et al.}

\begin{document}

\title{Discovery of a Little Red Dot candidate at $z\gtrsim10$ in COSMOS-Web based on MIRI-NIRCam selection}

\suppressAffiliations 

\input{author}

\collaboration{99}{}
\collaboration{0}{(Affiliations can be found after the references)}

\begin{abstract} 
JWST has revealed a new high-redshift population called little red dots (LRDs).
Since LRDs may be in the early phase of black hole growth, identifying them in the early universe is crucial for understanding the formation of the first supermassive black holes.
However, no robust LRD candidates have been identified at $z>10$, because commonly-used NIRCam photometry covers wavelengths up to $\sim5\,{\rm \mu m}$ and is insufficient to capture the characteristic V-shaped spectral energy distributions (SEDs) of LRDs.
In this study, we present the first search for $z\gtrsim10$ LRD candidates using both NIRCam and MIRI imaging from COSMOS-Web, which provides the largest joint NIRCam-MIRI coverage to date ($0.20\,{\rm deg^2}$). 
Taking advantage of MIRI/F770W to remove contaminants, we identify one robust candidate, CW-LRD-z10 at $z_{\rm phot}=10.5^{+0.7}_{-0.6}$ with $M_{\rm UV}=-19.9^{+0.1}_{-0.2}\,{\rm mag}$.
CW-LRD-z10 exhibits a compact morphology, a distinct V-shaped SED, and a non-detection in F115W, all consistent with being an LRD at $z\sim10$. 
Based on this discovery, we place the first constraint on the number density of LRDs at $z\sim10$ with $M_{\rm UV}\sim-20$ of $1.2^{+2.7}_{-1.0}\times10^{-6}\,{\rm Mpc^{-3}\,mag^{-1}}$, suggesting that the fraction of LRDs among the overall galaxy population increases with redshift, reaching $\sim3\%$ at $z\sim10$.
Although deep spectroscopy is necessary to confirm the redshift and the nature of CW-LRD-z10, our results imply that LRDs may be a common population at $z>10$, playing a key role in the first supermassive black hole formation.
\end{abstract}


\section{Introduction}\label{s:intro}
The James Webb Space Telescope (JWST, \citealt{Rigby2023_JWST}) has revolutionized the study of high-redshift active galactic nuclei (AGNs).
Despite its limited field-of-view, JWST's deep imaging has led to the discovery of numerous AGNs with the UV absolute magnitude of $M_{\rm UV}\gtrsim-22$\footnote{In this study, we define $M_{\rm UV}$ as the absolute magnitude in rest-frame 1450\,{\AA}, following the definition in \cite{Kocevski2024}}, fainter than typical UV luminous quasars with $M_{\rm UV}\lesssim-24$, even at $z\gtrsim4$ \citep[e.g.,][]{Onoue2023, Kocevski2023, Harikane2023_agn, Maiolino2023, Scholtz2023, Taylor2025_smbh}.
Assuming a single-epoch virial mass estimator \citep[e.g.,][]{Laor1998, Vestergaard2006, Shen2013} with broad lines, these AGNs are found to host supermassive black holes (SMBHs) with black hole masses of \hbox{$M_{\rm BH}\gtrsim10^6 M_\odot$}, reaching to $\sim3$\,dex lower than the masses of the most luminous quasars at similar epochs \citep[e.g.,][]{Fan2001, Willott2010, Shen2011, Wang2021, Yang2023_quasar, Banados2023}.
Recent studies have reported that the luminosity function of these JWST-found AGNs at the high-$z$ Universe \citep[e.g.,][]{Kokorev2024_census, Kocevski2024, Akins2024, Lin2025_c3d} is $1\,\mathchar`-\,2$\,dex higher than the extrapolation of the luminosity function of UV luminous quasars based on the ground-based surveys \citep{Matsuoka2016, Akiyama2018, Niida2020, Shen2020}, suggesting the rapid formation of an abundant SMBH population in the early Universe.

In particular, a previously unseen population that is likely associated with SMBHs has emerged, namely Little Red Dots (LRDs), from early JWST observations \citep[e.g.,][]{Labbe2023a, Furtak2024, Matthee2024, Kocevski2023, Akins2023, Barro2024, Akins2024, Greene2024, Setton2024, Hainline2025, Hviding2025}.
LRDs are characterized by their compact morphology in the rest-optical wavelength, with the strongest constraint of $r_e \lesssim 30\,{\rm pc}$ provided by \cite{Furtak2024} through strong gravitational lensing.
LRDs also exhibit unique V-shaped spectral energy distributions (SEDs), which consist of a rest-optical steep red continuum and a rest-UV excess \citep[e.g.,][]{Labbe2023a, Furtak2024, Matthee2024, Kocevski2023, Akins2023, Barro2024, Akins2024} with a pronounced slope change around the Balmer limit \citep[e.g.,][]{Setton2024}, often with a Balmer break \citep[e.g.,][]{Kokorev2024_break, Labbe2024, Wang2024a, deGraaff2025, Naidu2025}.
Spectroscopic observations have found broad ($>1000\,{\rm km\,s^{-1}}$) hydrogen Balmer emission lines in photometrically selected LRD candidates.
Assuming these broad lines are attributed to type-I AGNs and applying the single-epoch virial mass estimator \citep[e.g.,][]{Laor1998, Vestergaard2006, Shen2013}, previous studies have estimated their black hole masses as $M_{\rm BH}\gtrsim10^6 M_\odot$ \citep[e.g.,][]{Labbe2023b, Furtak2024, Matthee2024, Greene2024}.

Unlike typical low-redshift AGNs, LRDs remain undetected in X-rays \citep{Yue2024, Ananna2024, Maiolino2024_chandra, Akins2024, Lin2025_lowz} and radio \citep{Akins2024, Perger2024, mazzolari2024, Lin2025_lowz}, even with stacking analyses.
While we cannot detect a significant variability in broadband photometry for most cases \citep{Kokubo2024, Zhang2024_variability, tee2024_variability, Zhou2025_variability}, some studies have reported significant detection of variability in the H$\alpha$ equivalent width (EW) thanks to longer time baselines due to gravitational lensing \citep{Furtak2025_variability, Ji2025_local}.
These non-detections of typical AGN-like features may suggest a non-AGN nature.
An explanation of LRDs not related to SMBHs is that they are massive, compact galaxies \citep[e.g.,][]{Labbe2023a, Akins2023, Akins2024}.
However, this scenario implies large stellar masses, which may exceed the number densities of such massive halos predicted by the $\Lambda$CDM model \citep{Boylan_Kolchin2023}, and the densest stellar system discovered so far \citep{Baggen2024, Labbe2024, Ma2025_uncover}. 
On the other hand, it has also been proposed that super-Eddington accretion could potentially explain the absence of X-ray and variability signatures in cases where LRDs are indeed AGNs \citep[e.g.,][]{Pacucci2024_lrd, Inayoshi2024, Lambrides2024, Madau2024_super_eddington}.

One of the surprising aspects of LRDs is their abundance.
The number density of LRDs peaks at $z\sim6\,\mathchar`-\,8$, corresponding to about 3\% and 10\% of the galaxy population with $M_{\rm UV}=-20$ and $-22$, respectively, at $z\sim7$ \citep{Kocevski2024}.
It has also been reported that LRDs are more abundant by $\sim0.6\,{\rm dex}$ at $z\sim5$ compared to X-ray AGNs \citep{Parsa2018} and by $\sim1\,{\rm dex}$ at $z\sim7$ compared to color-selected AGNs \citep{Kulkarni2019} with the same UV luminosity \citep{Kocevski2024}, suggesting that LRDs represent a crucial population for understanding BH evolution in the early universe.

Typical photometric selection for LRD candidates is performed using their compactness and characteristic V-shape SEDs in JWST/NIRCam photometry.
For instance, \cite{Labbe2023a} and \cite{Kokorev2024_census} used multiple color-selection criteria, while \cite{Kocevski2024} used SED slopes to identify V-shape profiles.
Similarly, \cite{Akins2024} employed the single color threshold of $m_{\rm F277W} - m_{\rm F444W} > 1.5$ and compactness criterion to select LRD candidates.

Spectroscopic observations of these photometrically selected LRD candidates have detected broad Balmer emission lines, confirming their nature as LRDs \citep[e.g.,][]{Greene2024}.
Spectroscopic surveys have confirmed that these photometrically selected LRD candidates are highly reliable, with low contamination rates ($\sim 90\%$).
However, they also suggested a completeness of $\sim50\%$, indicating that a part of the LRD population may have been missed in previous photometric selections \citep{Lin2025_c3d, Hviding2025}.
Among previously selected LRD candidates, some studies \citep[e.g.,][]{Hainline2025, Zhang2025_nl} have reported contamination where a red continuum could not be spectroscopically confirmed, and the V-shaped SED seen in the photometry was instead caused by strong emission lines.
However, the fraction of such contamination cases is estimated to be $\sim20\,\mathchar`-\,35\%$ of LRD candidates, implying that the majority of candidates exhibit both a red continuum and broad Balmer emission lines in their spectrum.

Based on the redshift distribution from \cite{Kocevski2024}, \cite{Inayoshi2025} reported that the number density of LRDs with $M_{\rm UV}<-18$ follows a log-normal distribution as a function of cosmic time, suggesting that LRDs may be driven by stochastic phenomena.
With LRDs seemingly being a distinct population with unique characteristics in the early Universe, one hypothesis is that they may represent the first or second accretion episodes following the formation of their seed black holes, potentially undergoing rapid growth via super-Eddington accretion \citep{Inayoshi2025}.
To test the evolution of such an LRD population observationally, we need to constrain the LRD population over a wide redshift range.
Efforts to identify LRDs at lower redshifts ($z < 4$, e.g., \citealt{LinR2025_lowzGP}) are not limited to JWST observations \citep{Kocevski2024, Zhuang2025}, but have also been extended to wider-area surveys using Euclid (\!\!\citealt{Euclid2025}) and Subaru's Hyper Suprime-Cam (HSC) imaging \citep{Ma2025}, combined with archival infrared survey data.
Recently, \citet{Lin2025_lowz} and \citet{Ji2025_local} identified LRD analogues even at $z \sim 0.1\,\mathchar`-\,0.2$, based on spectroscopically confirmed objects from the Sloan Digital Sky Survey (SDSS).
Their results are consistent with a decline from $z\sim5\,\mathchar`-\,6$ or indicate a higher number density than predicted by \cite{Inayoshi2025}; however, there is still a possibility of contamination since sources lack spectroscopic confirmation.

Expanding LRD searches not only toward lower redshift but also to higher redshift is crucial for constraining the evolution of the LRD population itself, as well as for probing the first SMBHs in the Universe.
Currently, three spectroscopically confirmed AGNs (not LRDs) have been reported at $z_{\rm spec}>10$: GHZ-9 ($z_{\rm spec} = 10.145 \pm 0.010$; \citealt{Kovacs2024_GHZ9, Napolitano2025_GHZ9}), UHZ-1 ($z_{\rm spec} = 10.073 \pm 0.002$; \citealt{Bogdan2024_UHZ1, Goulding2023}), and GN-z11 ($z_{\rm spec} = 10.6034 \pm 0.0013$; \citealt{Oesch2016, Jiang2021_GNz11, Bunker2023, Maiolino2023_GNz11, Scholtz2024_Gnz11}).
Whereas, the current highest spectroscopic redshift for an LRD is CAPERS-LRD-z9 at $z_{\rm spec}=9.288$ \citep{Taylor2025}, which was selected through photometry \citep{Kocevski2024, Akins2024, Barro2024_comprehensive}.

A previous study \citep{Kocevski2024} selected 13 LRD candidates at $z_{\rm photo}\gtrsim10$ solely from NIRCam photometry.
Among these, six objects have been spectroscopically observed; however, no clear LRDs have been confirmed at $z\gtrsim10$ (see Appendix\,\ref{Ap:Kocevski} for each case).
As suggested in \cite{Kocevski2024} and \cite{Hainline2025}, this demonstrates that selecting $z\sim10$ LRDs solely with NIRCam is inherently challenging.
Since F444W, the longest wavelength filter in NIRCam, is too close to the Balmer limit at $z\sim10$ ($\sim4\,{\rm \mu m}$), it is difficult to separate the contribution of emission lines or a stellar-origin Balmer break from the characteristic red slope of LRDs.
With MIRI photometry at longer wavelengths, it becomes possible to remove such contamination and select $z\gtrsim10$ LRDs more reliably.
In this context, \cite{PerezGonzalez2024_nircamdark} reported a NIRCam dark galaxy detected in MIRI/F1000W.
While they discuss the possibility that this object is an LRD at $z \sim 14$, it is fainter than $30\,\mathchar`-\,30.5$ mag in the NIRCam wavelength range and lacks spectroscopic confirmation; thus, its true nature remains uncertain.
Based on the log-normal distribution proposed by \cite{Inayoshi2025}, the number density of LRDs declines toward higher redshifts from $z\sim6$.
Thus, to robustly identify $z\sim10$ LRDs, a large survey volume with joint NIRCam and MIRI coverage is essential.

In this study, we utilize data from COSMOS-Web, which provides the largest combined NIRCam and MIRI survey area to date ($\sim0.18\,{\rm deg^2}$), to search for $z\sim10$ LRDs and for the first time place constraints on their luminosity function.
In Section\,\ref{s:data} we describe the data, and in Sections\,\ref{s:methods} and \ref{s:analysis}, we present the selection methods and the follow-up analysis of a selected candidate.
Section\,\ref{s:results} shows the results of the follow-up analysis of the candidate.
Then, in Section\,\ref{s:discussion}, we constrain the luminosity function of LRDs at $z\sim10$ and discuss their implications.
Throughout this paper, we use the AB magnitude system \citep{Oke1983} and assume a standard cosmology with $H_0 = 70~{\rm km~s^{-1}~Mpc^{-1}}$, $\Omega_m = 0.30$, and $\Omega_\Lambda=0.70$.

\begin{figure*}[ht!]
\epsscale{1.}
\plotone{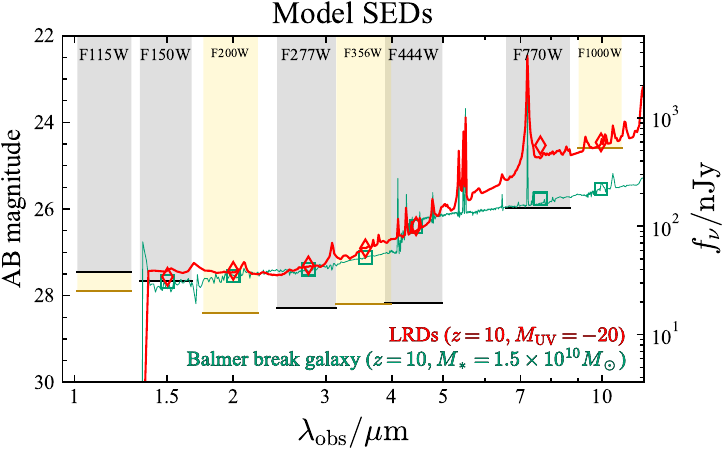}
\caption{Model SED of a $z=10$ LRD with $M_{\rm UV} = -20$ (red) and $5\sigma$ depth for point sources in COSMOS-Web (F115W, F150W, F277W, F444W, and F770W) from \citet{Casey2023} and in COSMOS-3D (F115W, F200W, F356W, and F1000W) shown in gray and gold rectangles, respectively.
Model SEDs of potential contaminants of the $z\sim10$ LRD photometric selection, a $z=10$ galaxy with a stellar-origin Balmer break with $A_V=0.5$ and $M_* = 1.5 \times 10^{10}\,M_\odot$, is overlaid.
The LRD model SED is from \cite{Akins2024}, and the galaxy SED is generated using FSPS \citep{Conroy2009, Conroy2010}.
The predicted model photometry is shown with diamonds for the LRD and squares for the galaxy.
MIRI photometry is essential to distinguish between $z\sim10$ LRDs and $z\sim10$ galaxies with stellar-origin Balmer break.
\label{fig:depth}}
\end{figure*}

\section{Data}\label{s:data}

\subsection{Survey field}\label{ss:survey-field}
As discussed above, we require data from a wide-area survey that combines JWST/NIRCam \citep{Rieke2023} and MIRI \citep{Bouchet2015} imaging to robustly search for $z\sim10$ LRDs.
Therefore, we use COSMOS-Web (\href{https://www.stsci.edu/jwst/science-execution/program-information?id=1727}{GO\,1727}, PI: J. Kartaltepe and C. Casey, \citealt{Casey2023}), a JWST Cycle 1 Treasury Program with the largest NIRCam and MIRI coverage to date.
COSMOS-Web covers $0.54\,{\rm deg^2}$ with NIRCam using four filters: F115W, F150W, F277W, and F444W.
Simultaneously, COSMOS-Web covers $0.20\,{\rm deg^2}$ with MIRI F770W.
The overlapping area between the NIRCam and MIRI coverage is $0.18\,{\rm deg^2}$, which is the effective field for our LRD search utilizing both NIRCam and MIRI photometry.
Figure\,\ref{fig:depth} illustrates the depth of COSMOS-Web observations with a stacked model SED of photometrically-selected LRDs provided in \cite{Akins2024}\footnote{\url{https://github.com/hollisakins/akins24_cw/}}, assuming $z=10$ and $M_{\rm UV}=-20$ (red line).
To separate LRDs from other lower-$z$ dusty sources based on the presence of a UV-excess with F150W and F277W, we focus only on sources brighter than $M_{\rm UV}\lesssim-19.5$ mag, which can be detected in both F150W and F277W with a $\gtrsim3\sigma$ significance.

The COSMOS-Web field is also partially covered by COSMOS-3D (\href{https://www.stsci.edu/jwst-program-info/program/?program=5893}{GO\,5893}, PI: K. Kakiichi), which includes NIRCam imaging in F115W, F200W, and F356W. Observations are ongoing and COSMOS-3D will cover a part of the full COSMOS-Web field: $\sim 0.33\,{\rm deg^2}$ ($\sim 0.14\,{\rm deg^2}$) with NIRCam (MIRI).
To uniformly select $z\sim10$ LRDs across the wider field, the color selection in this study only utilizes photometry from COSMOS-Web.

\subsection{Imaging data}\label{ss:image_data}
The candidate selection process (Section\,\ref{s:methods}) utilizes the catalog photometry from COSMOS2025 (\citealt{Shuntov2025_COSMOS2025}, Section\,\ref{ss:catalog}): JWST/NIRCam (F115W, F150W, F277W, and F444W) and JWST/MIRI (F770W) from COSMOS-Web combined with Subaru/HSC ($ugriz$) and HST/ACS (F814W).
For the subsequent follow-up analysis of the selected candidate (Section\,\ref{s:analysis}), we utilize all available imaging data, including JWST/NIRCam (F115W, F200W, and F356W) and JWST/MIRI (F1000W and F2100W) from COSMOS-3D and HST/ACS (F606W).

\subsubsection{JWST/NIRCam}
We reduce all available NIRCam imaging data through a custom pipeline as described in \cite{Franco2025_CW_NIRCam} (see Akins in prep. for details).
The raw NIRCam images are processed through the JWST Calibration Pipeline version 1.17.1 \citep{jwstpipe_1171}, with the addition of several custom modifications, including the subtraction of $1/f$ noise and sky background.
We use the Calibration Reference Data System (CRDS)\footnote{\url{jwst-crds.stsci.edu}} pmap 1331.
The $1/f$ noise is subtracted through the iterative source masking and amp-row median subtraction as described in \cite{Bagley2023}, and wisps---scattered light features present in several NIRCam short wavelength (SW) detectors---are removed by rescaling and subtracting the version 3 templates provided by STScI\footnote{\url{https://stsci.app.box.com/s/1bymvf1lkrqbdn9rnkluzqk30e8o2bne}}. 
Background subtraction is performed using the iterative source masking procedure described in \citet{Bagley2023}. 
Astrometric calibration is conducted via the \textit{JWST}/\textit{HST} alignment tool \citep[JHAT;][]{Rest2023}, with a reference catalog based on an \textit{HST}/F814W $0.\!\!^{\prime\prime}03\,{\rm pixel^{-1}}$ mosaic of the field \citep{Koekemoer2007} with astrometry tied to Gaia-DR3 (\!\!\citealt{GAIA_DR3}).
All NIRCam mosaics are resampled to a uniform pixel scale of $0.\!\!^{\prime\prime}03\,{\rm pixel^{-1}}$.
The $5\sigma$ point source limiting magnitude for COSMOS-Web NIRCam images is summarized in \cite{Franco2025_CW_NIRCam}, and those for COSMOS-3D images are 28.4 and 28.2 in F200W and F356W, respectively.
Note that F115W has been processed by combining both the COSMOS-Web and COSMOS-3D data, resulting in a point-source limiting magnitude of 27.9.

\subsubsection{JWST/MIRI}
For MIRI imaging, we use F1000W and F2100W imaging from COSMOS-3D with COSMOS-Web's F770W, thereby extending our MIRI wavelength coverage up to $\sim 25\,{\rm \mu m}$. 
The F770W data reduction of COSMOS-Web is presented in \cite{Harish2025_CW_MIRI}, with key steps briefly summarized below.
The F770W data were reduced using the JWST Calibration Pipeline version 1.12.5 \citep{jwstpipe_1125} with JWST Calibration Reference Data System Pipeline Context version 1130.
The reduction includes the standard rate calculation, jump detection, slope fitting, non-linearity correction, flat fielding, flux scale conversion, etc., as well as a highly customized master background subtraction (see \citealt{Harish2025_CW_MIRI} for the demonstration of quality improvement).
The astrometry correction is also highly customized with source extraction and aligning using SExtractor and JWST pipeline's \texttt{tweakreg} module, and with the same absolute reference catalog as for NIRCam reduction.
The Lyot detector area was also processed together with the main imager area to increase the field of view, which provides good quality of data after applying our customized master background subtraction.
F770W image is also resampled to a pixel scale of $0.\!\!^{\prime\prime}03\,{\rm pixel^{-1}}$.
The F1000W and F2100W data were reduced using the JWST Calibration Pipeline version 1.12.5 \citep{jwstpipe_1125} with JWST Calibration Reference Data System Pipeline Context version 1303.
F1000W and F2100W images are resampled to a pixel scale of $0.\!\!^{\prime\prime}11\,{\rm pixel^{-1}}$.

\subsubsection{HST/ACS}
For the HST data, we make use of the ACS F814W mosaics \citep{Koekemoer2007}, which have been reprocessed with up-to-date calibration reference files and aligned to Gaia-DR3 (\!\!\citealt{GAIA_DR3}), serving as the astrometric reference frame for all the above JWST data.
We also make use of the ACS F606W mosaics, which are a full-depth combination of data from CANDELS \citep{Grogin2011, Koekemoer2011} and CLUTCH (HST program ID 17802, PI: J. Kartaltepe), also aligned to the HST ACS F814W astrometric grid.
These HST mosaics have been resampled to a uniform pixel scale of $0.\!\!^{\prime\prime}03\,{\rm pixel^{-1}}$.

\subsection{Photometric catalog}\label{ss:catalog}
We select LRD candidates using photometry from the COSMOS2025 catalog \citep{Shuntov2025_COSMOS2025}, which provides a comprehensive photometric dataset for the COSMOS-Web field containing 37 bands covering $0.3\,\mathchar`-\,8\,{\rm \mu m}$ from Subaru/HSC to JWST/MIRI (not including COSMOS-3D photometry).
Among them, we use the information from Subaru/HSC ($g$, $r$, $i$, $z$, and $y$), HST/ACS (F814W), JWST/NIRCam (F115W, F150W, F277W, F444W), and JWST/MIRI (F770W) in the selection process (Section\,\ref{ss:color_selection}).
Note that we use the photometry from COSMOS2025 only for the photometric selection described in Section\,\ref{ss:color_selection}.
After the selection, we remeasure the photometry (Section~\ref{ss:forced_photo}) based on the imaging data described in Section\,\ref{ss:image_data} to apply a consistent method across all available bands, including filters not covered in the COSMOS2025 catalog.

\subsection{Grism data}\label{ss:grism}
COSMOS-3D also performs a Wide Field Slitless Spectroscopy (WFSS) survey with NIRCam/F444W, covering a part of the COSMOS-Web field.
This WFSS observation has covered the selected candidate described later.
The COSMOS-3D grism data is processed with two different processing software, \texttt{unfold\_jwst} \citep{Wang2023} and \texttt{Grizli}\footnote{\url{https://github.com/gbrammer/grizli}} \citep{grizli_2023}, to reduce the software dependencies and obtain robust results (see K. Kakiichi in prep. for detailed data processing).

The candidate has also been covered in the NIRCam/F322W2 WFSS observation in POPPIES (\href{https://www.stsci.edu/jwst-program-info/program/?program=5398}{GO\,5398}, PI: J. Kartaltepe and M. Rafelski).
However, CW-LRD-z10 lies near the edge of the WFSS field of view, and a usable spectrum could not be obtained in POPPIES.

\section{Sample selection}\label{s:methods}
\subsection{Evolution on the color-color diagram}\label{ss:color-color}
\begin{figure*}[ht!]
\epsscale{1.15}
\plotone{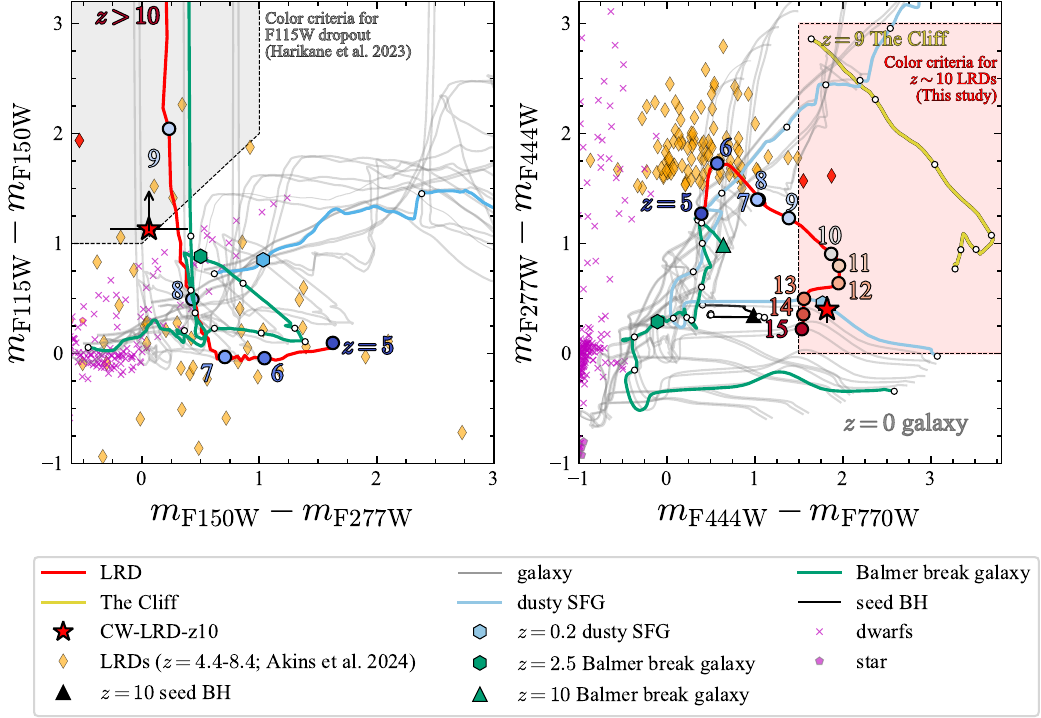}
\caption{
Color-color diagrams based on the LRD model SEDs from \cite{Akins2024} (red lines).
The left and right panels show $m_{\rm F115W} - m_{\rm F150W}$ vs. $m_{\rm F150W} - m_{\rm F277W}$ and $m_{\rm F277W} - m_{\rm F444W}$ vs. $m_{\rm F444W} - m_{\rm F770W}$, respectively. 
The red filled region defined by $0 < m_{\rm F277W} - m_{\rm F444W} < 2.25$ and $1.5 < m_{\rm F444W} - m_{\rm F770W}$ can effectively distinguish $z \sim 10$ LRDs from other objects.
The selected candidate in this study, CW-LRD-z10, is highlighted with a red star with error bars.
Note that $m_{\rm F115W}-m_{\rm F150W}$ color for CW-LRD-z10 is a $2\sigma$ lower limit. 
CW-LRD-z10 also passes the color criteria for F115W dropouts \citep{Harikane2023_highz} shown by the gray shaded region in the left panel.
Model galaxy SEDs generated using FSPS \citep{Conroy2009, Conroy2010} with various star formation histories and dust attenuation values at $z=0\,\mathchar`-\,10$ are shown in gray solid lines. 
Of these, we highlight two cases: (1) a stellar origin Balmer break (with an $f_\nu$ ratio of $\sim2$) and $A_V=0.5$ (green), and (2) a dusty SFG model with $A_V=5$ (cyan).
Seed BH models from \cite{Inayoshi2022} at $z=8\,\mathchar`-\,14$ and LRDs with an extremely strong Balmer break at $z=9\,\mathchar`-\,15$, assuming the spectrum of The Cliff in \cite{deGraaff2025}, are shown in black and orange, respectively.
For these model SEDs, dots are placed at redshift intervals of $\Delta z=1$.
Specifically, $z=10$ Balmer break galaxy, $z=10$ seed BH, $z=2.5$ galaxy with stellar-origin Balmer break, and $z=0.2$ dusty SFG are shown in green triangle, black triangle, green hexagon, and cyan hexagon, respectively.
Photometrically-selected LRDs from \citet{Akins2024} with $S/N > 3$ are shown as orange diamonds, among them are two objects that pass our color selection (red diamonds).
Dwarf stars and main sequence stars from the BT-Settl model \citep{Allard2012} are plotted in magenta crosses and pentagons, respectively.
\label{fig:cc}}
\end{figure*}

To define robust color thresholds for selecting $z\sim10$ LRDs, we model their redshift evolution on a color-color diagram.
The thick red line in Figure\,\ref{fig:cc} illustrates the redshift evolution of LRDs on the $m_{\rm F115W}-m_{\rm F150W}$ vs. $m_{\rm F150W}-m_{\rm F277W}$ and the $m_{\rm F277W}-m_{\rm F444W}$ vs. $m_{\rm F444W}-m_{\rm F770W}$ planes.
Here, we assume the model SED for stacked LRDs from \cite{Akins2024}, evolved from $z\sim5$ to $15$.

We also plot the color evolution of LRDs with a strong Balmer break by using the NIRSpec/PRISM spectrum of such an LRD at $z=3.548$, ``The Cliff''  \citep{deGraaff2025}, obtained in \href{https://www.stsci.edu/jwst/science-execution/program-information?id=4233}{GO\,4233} (RUBIES, PI: A. de Graaff and G. Brammer).
We download the spectrum from the Dawn JWST archive\footnote{\url{https://dawn-cph.github.io/dja/}}, which was processed with \texttt{msaexp} \citep{msaexp, Heintz2024, deGraaff2024_msa}, a tool for extracting spectra directly from the telescope exposures.
We further perform flux calibration to correct for possible slit loss by adopting a 2${}^{\rm{nd}}$-order polynomial function to match the spectrum with the NIRCam photometry (F090W, F115W, F150W, F200W, F277W, F356W, F444W), taken from the DJA morphological catalogs based on \texttt{SourceExtractor++} \citep{Bertin_sepp, Kummel_sepp}.
The correction factor ranges from 0.96 to 1.08, indicating that the DJA spectrum and photometry for The Cliff are basically consistent with each other (differences are $<0.1$\,mag).
Note that The Cliff is not plotted in Figure\,\ref{fig:cc} (left) because we cannot estimate its F115W and F150W photometry at these redshifts due to the limited wavelength coverage of the spectral data ($\lambda_{\rm rest}\gtrsim0.14\,{\rm \mu m}$).
The stacked LRD SED \citep{Akins2024} has a rest-frame H$\alpha$ EW of $\sim700\,{\rm \AA}$ (including both broad and narrow components) and the rest-frame optical continuum slope, measured in $0.38<\lambda_{\rm rest}/{\rm\mu m}<0.8$ after masking H$\beta$+[O\,{\sc iii}] and H$\alpha$ lines, of $\beta\sim0.4$ ($f_\lambda\propto\lambda^\beta$).
The Cliff has a lower rest-frame H$\alpha$ EW of $\sim400\,{\rm \AA}$ and a similar optical continuum slope of $\beta\sim0.3$.
Thus, The Cliff has an H$\alpha$ contribution and an optical slope similar to the stacked model SED (though slightly weaker H$\alpha$), with the main difference being its much stronger Balmer break.

To assess the possibility of contamination, in Figure\,\ref{fig:cc}, we also plot the color evolution of the predicted SED of seed BHs that assumes a super-Eddington accreting SMBHs with $M_{\rm BH}\sim10^6M_\odot$ (\citealt{Inayoshi2022}, $z=8\,\mathchar`-\,14$, black line) and Galactic stars (main-sequence and dwarfs assuming $T_{\rm eff}=400\,\mathchar`-\,4000\,{\rm K}$) from the BT-Settl model with solar metallicity (\citealt{Allard2012}, magenta pentagon and cross, respectively).
Additionally, we generate galaxy SED models using Flexible Stellar Population Synthesis (\texttt{FSPS}, \citealt{Conroy2009, Conroy2010}), assuming a delayed-$\tau$ star formation history with $t_{\rm age}=11.3\,{\rm Gyr}$, which is 0.95 times of the Hubble time at $z=0.2$, $\tau=0.01$, $0.1$, $1$, and $10\,{\rm Gyr}$, and dust attenuation with $A_V=0$, $0.5$, $1$, $3$, and $5$ mags.
These galaxy SED models include nebular emission lines \citep{Byler2017} based on \texttt{Cloudy} \citep{CLOUDY90, cloudy2017}. We fix the gas ionization parameter to $\log U=-2$ and gas metallicity to $\log\left(Z_{\rm gas}/Z_\odot\right)=-1$. We indicate the color evolution of the synthetic galaxies at $z=0\,\mathchar`-\,10$ (gray lines).

From Figure\,\ref{fig:cc} (left), we see that LRDs have a high value for the $m_{\rm F115W} - m_{\rm F150W}$ color due to the Lyman break entering F115W, a feature similar to galaxies at $z>8$.
However, the $m_{\rm F150W} - m_{\rm F277W}$ color also overlaps significantly with other galaxy populations. 
Therefore, the $m_{\rm F115W} - m_{\rm F150W}$ and $m_{\rm F150W} - m_{\rm F277W}$ colors are insufficient to distinguish $z\sim10$ LRDs from normal $z\sim10$ galaxies, while being effective at identifying dropouts as shown by the color criteria for F115W dropout galaxies (gray-filled region, from \citealt{Harikane2023_highz}).

From Figure\,\ref{fig:cc} (right), LRDs have $m_{\rm F277W} - m_{\rm F444W}>1.5$ at $z\sim6$, which is the selection threshold for LRDs in \cite{Akins2024}.
Toward higher redshift, the Balmer limit shifts closer to F444W, i.e., F277W traces the flat rest-UV part of the spectrum, and H$\alpha$ and [O{\sc iii}]+H$\beta$ move out of the F444W coverage.
As a result, the $m_{\rm F277W} - m_{\rm F444W}$ color decreases to $\sim1.0$ at $z\sim10$.
Thus, the $m_{\rm F277W} - m_{\rm F444W}>1.5$ color cut used in \cite{Akins2024} is not effective for $z\sim10$ LRDs. Because the Balmer limit shifts to $\sim4\,{\rm \mu m}$, and H$\alpha$ falls into F770W at $z\sim10$, the $m_{\rm F444W} - m_{\rm F770W}$ color increases, reaching $m_{\rm F444W} - m_{\rm F770W}\sim2$ at $z\sim10$.
We also note that The Cliff exhibits a larger color, driven by its stronger Balmer break compared to stellar-origin Balmer breaks.

As shown, the color space, $m_{\rm F277W} - m_{\rm F444W}$ vs. $m_{\rm F444W} - m_{\rm F770W}$, effectively isolates $z\sim10$ LRDs from other sources. 
Seed BHs, showing strong H$\alpha$ with rest-frame EW of $\sim1000$\,\AA, have $m_{\rm F444W} - m_{\rm F770W}<1.5$ due to the absence of a red continuum spectrum.
Brown dwarfs and main-sequence stars exhibit $m_{\rm F444W} - m_{\rm F770W}\lesssim0$, far from $z\sim10$ LRDs.
Most normal galaxies are also well separated from $z\sim10$ LRDs, though dusty star-forming galaxies (SFG) at $z\sim0.2$ could mimic LRD-like $m_{\rm F277W} - m_{\rm F444W}$ and $m_{\rm F444W} - m_{\rm F770W}$ colors.
At $z\sim0.2$, polycyclic aromatic hydrocarbon (PAH) emission bands at the rest-frame $3.3\,{\rm \mu m}$ and $6.2\,{\rm \mu m}$ enter F444W and F770W, respectively, boosting the observed fluxes in those bands.
Thus, $z\sim0.2$ dusty SFGs can also reproduce an apparent slope change around $\lambda_{\rm obs} \sim 3\,\mathchar`-\,4\,{\rm \mu m}$, resembling the photometric SED of $z\sim10$ LRDs.
In Section\,\ref{s:results}, we further discuss the possibility of interlopers based on their SEDs and morphology.

\subsection{Selection procedures}\label{ss:color_selection}
We perform color-based selection using the COSMOS2025 photometric catalog (Section\,\ref{ss:catalog}).
The number of objects at each step in the selection is summarized in Table\,\ref{tab:numbers}.
The COSMOS2025 catalog contains 784,016 objects, of which 694,631 have \texttt{warn\_flag=0} (good objects), indicating that they are not hot pixels or artifacts and there is no inconsistency between ground and space observations (see \citealt{Shuntov2025_COSMOS2025} for detail).
Among them, 236,127 objects are located in the area with MIRI coverage and constitute the parent sample used in this study.

First, we require a detection with $S/N>5$ in F150W, F277W, F444W, and F770W, while undetected ($S/N<2$) in F814W and F115W using the {\tt snr\_\{filter\}} columns, which give $S/N$ within $0.\!\!^{\prime\prime}2$ diameter apertures.
As shown in Figure\,\ref{fig:depth}, the shallowest photometry relative to $z\sim10$ LRD's SED in our dataset, i.e., the bottleneck for identifying $z\sim10$ LRDs, is F150W.
Sources with $S/N_{\rm F150W}\sim5$ in COSMOS2025 has the median F150W photometry of 27.7\,mag, which corresponds to $M_{\rm UV}\sim-19.5$ at $z\sim9$.
Therefore, LRDs with $M_{\rm UV}\gtrsim-19.5$ are not detected in F150W and thus cannot be selected in our sample.
To exclude low-redshift galaxies, we also apply additional non-detection ($S/N<2$) criteria in all of the HSC $g$, $r$, $i$, $z$, and $y$ bands ({\tt flux\_model\_hsc-\{filter\}}, {\tt flux\_err-cal\_model\_hsc-\{filter\}} columns).

We adopt the following new color criteria
\begin{align}
    m_{\rm F277W} - m_{\rm F444W} &> 0, \label{eq:col1}\\
    m_{\rm F277W} - m_{\rm F444W} &< 3.0, \label{eq:col2}\\
    m_{\rm F444W} - m_{\rm F770W} &> 1.5, \label{eq:col3}
\end{align}
as shown in the red-filled region in the color-color diagram shown in the right panel of Figure\,\ref{fig:cc}.
Magnitudes are from the columns {\tt mag\_auto\_\{filter\}}, measured in small elliptical apertures with corrections for the Kron aperture and PSF differences.
Note that this color selection is capable of identifying $z\gtrsim9$ LRDs even in cases where they exhibit an extremely strong Balmer break, such as seen in The Cliff \citep{deGraaff2025}.

To select compact sources, we compare aperture photometry measured in $0.\!\!^{\prime\prime}2$ and $0.\!\!^{\prime\prime}5$ apertures on F444W images:
\begin{equation}
    {\rm compactness} = \frac{f_{\rm F444W}\left(0.\!\!^{\prime\prime}2\right)}{f_{\rm F444W}\left(0.\!\!^{\prime\prime}5\right)}.
\end{equation}
Then, we require the compactness to be between 0.5 and 0.7 following \cite{Akins2024}, where the lower and upper limits exclude extended objects and imaging artifacts, respectively.
The corresponding rest-frame wavelength of F444W at $z\sim10$ ($\sim0.4\,{\rm \mu m}$) is shorter than at $z\sim6$ ($\sim0.6\,{\rm \mu m}$), which is the median photometric redshift of the sample in \citet{Akins2024}.
However, F444W still probes the rest-frame optical regime, longer than the Balmer limit.
Additionally, F770W has lower spatial resolution compared to F444W.
Therefore, we continue to use F444W for compactness measurements.
We also confirm in Section\,\ref{ss:morph} that the selected source is compact in other JWST filters.

We identify six objects after applying the aforementioned selection based on $S/N$, color, and compactness. Visual inspection of their images reveals that four of these are indistinguishable from noise or artifact patterns in F770W.
Another object shows tentative detections in both the HSC and HST F435W images. Therefore, five objects are excluded with their cutout images shown in Appendix\,\ref{ap:visual_inspection}.

As a result, we find one LRD candidate at $z\sim10$ (CW-LRD-z10; ID: \texttt{426069} in COSMOS2025).
The images and SED of this object are shown in Figure~\ref{fig:sedfit}, and its key information is summarized in Table~\ref{tab:info}.

Since compactness can be underestimated in LRDs with companions \citep{Tanaka2024_dLRD}, we also implement a relaxed compactness interval $0.4\,\mathchar`-\,0.7$, which provides no additional candidates.
Note that \texttt{warn\_flag = 2} and \texttt{3} in COSMOS2025 indicate cases where UVISTA fluxes significantly exceed the fluxes in the nearest NIRCam bands and those that are expected to be detected in HSC or UVISTA based on their NIRCam photometry but are not detected, respectively.
These cases are primarily due to blending with nearby brighter objects in the UVISTA or HSC images \citep{Shuntov2025_COSMOS2025}.
We confirm that none of the objects with \texttt{warn\_flag = 2} or \texttt{3} are selected with our method, even if we do not apply a $S/N$ cut in the HSC bands.

\begin{deluxetable*}{lll}
\tablecaption{The number of objects selected in each procedure \label{tab:numbers}}
\tablewidth{30pt}
\tablehead{
\colhead{Step} & \colhead{Selection} & \colhead{\#}
}
\startdata
0 & COSMOS2025 & 784,016\\
1 & \texttt{warn\_flag=0} & 694,341\\
2 & In MIRI coverage & 236,127\\
3 & $S/N$ in HSC, HST, and JWST & 259\\
4 & colors & 18\\
5 & compact & 6\\
6 & visual inspection & 1\\
\enddata
\end{deluxetable*}

\subsection{Comparison with previous studies}
The following objects, previously selected as high-$z$ LRD candidates in the COSMOS-Web field, are not selected in our study for various reasons.
\begin{itemize}
    \item Among the sources selected in \citet{Akins2024} with the color threshold of $m_{\rm F277W} - m_{\rm F444W} > 1.5$, two objects, ID:\texttt{49687} and ID:\texttt{77439} (IDS are from COSMOS2025\footnote{ID:\texttt{73111} and ID:\texttt{88514}, respectively, in the \citealt{Akins2024} catalog.}), satisfy our color conditions (Equations\,\ref{eq:col1}, \ref{eq:col2}, and \ref{eq:col3}) as shown in red diamonds in Figure\,\ref{fig:cc}.
    ID:\texttt{49687} is detected in F115W ($S/N=5.4$) and in HSC ($S/N=4.6$ in the $g$ band), thus does not pass the $S/N$ cut and is likely a low-$z$ object.
    On ID:\texttt{77439}, while the photometric redshift ($z_{\rm photo}=8.5^{+3.0}_{-3.6}$), given in COSMOS2025 is consistent with $z\sim10$, it is barely detected in F150W ($S/N=2.1$), hence does not pass the $S/N$ cut.
    Even if ID:\texttt{77439} is a genuine $z\sim10$ LRD, 
    $M_{\rm UV}$ would be -18.5 (based on F150W photometry), much lower than the our targeted luminosity range; thus it does not affect our final discussion about the LRD luminosity function at $z\sim10$ (Section\,\ref{ss:LF}).
    \item CAPERS-LRD-z9 at $z_{\rm spec}=9.288$ \citep{Taylor2025} lies within PRIMER-COSMOS, which overlaps with the COSMOS-Web field and is deeper than COSMOS-Web by $\sim1$\,mag.
    However, CAPERS-LRD-z9 is out of the MIRI/F770W coverage; thus, it is not selected with our method.
    Also note that CAPERS-LRD-z9 has $M_{\rm UV} = -18.2$, which is fainter than our target luminosity range of $M_{\rm UV}\lesssim-19.5$ limited by the F150W depth in COSMOS-Web (Figure\,\ref{fig:depth} and Section\,\ref{ss:color_selection}).
    CAPERS-LRD-z9 with $m_{\rm F150W}=29.12$ is indeed detected in the F150W observation in PRIMER-COSMOS; however, it is not detected in COSMOS-Web with a shallower F150W depth than PRIMER-COSMOS.
    Therefore, even if it had been covered by F770W, it would have not passed the $S/N$ cut.
    \item \cite{Kocevski2024} selected four objects as $z_{\rm photo}>10$ LRD candidates using only NIRCam photometry from PRIMER-COSMOS (see Appendix\,\ref{Ap:Kocevski}).
    However, three of the four candidates, PRIMER-COS-49611, PRIMER-COS-70533 (CAPERS-LRD-z9 discussed above, \citealt{Taylor2025}), and PRIMER-COS-78729, are not in the parent sample due to the lack of the COSMOS-Web F770W coverage.
    Also note that PRIMER-COS-78729 is confirmed to be $z_{\rm spec} = 5.9$ from archival spectroscopic data (see Appendix\,\ref{Ap:Kocevski}).
    The remaining source, PRIMER-COS-113415, is faint ($M_{\rm UV} = -17.58$ if we assume $z_{\rm photo} = 10.69$) and is not detected in COSMOS2025.
\end{itemize}

\begin{deluxetable*}{ll}
\tablecaption{Properties of CW-LRD-z10 \label{tab:info}}
\tablewidth{30pt}
\tablehead{
\colhead{Property} & \colhead{Value}
}
\startdata
ID in COSMOS2025 & \texttt{426069}\\
R.A. & 150.0840984\,deg \\
 & ($10^{\rm h}\,00^{\rm m}\,20.\!\!^{\rm s}1836$) \\
Decl. & 2.453895\,deg \\
 & ($2^{\circ}\,27^{\prime}\,14.\!\!^{\prime\prime}022$) \\
 \hline
$f_{\rm F606W} / {\rm n Jy}$ & $<12$ \\
$f_{\rm F814W} / {\rm n Jy}$ & $<3.3$ \\
$f_{\rm F115W} / {\rm n Jy}$ & $<11$ \\
$f_{\rm F150W} / {\rm n Jy}$ & $30.5\pm8.9$ \\
$f_{\rm F200W} / {\rm n Jy}$ & $37.9\pm1.2$ \\
$f_{\rm F277W} / {\rm n Jy}$ & $32.3\pm2.9$ \\
$f_{\rm F356W} / {\rm n Jy}$ & $32.5\pm2.0$ \\
$f_{\rm F444W} / {\rm n Jy}$ & $46.8\pm3.4$ \\
$f_{\rm F770W} / {\rm n Jy}$ & $\left(2.51\pm0.14\right)\times10^2$ \\
$f_{\rm F1000W} / {\rm n Jy}$ & $<2.5\times10^2$ \\
$f_{\rm F2100W} / {\rm n Jy}$ & $<1.6\times10^3$ \\
\hline
$z_{\rm photo}$ & $10.50^{+0.72}_{-0.58}$ \\
$M_{\rm UV}$ & $-19.9^{+0.1}_{-0.2}$\\
\enddata
\tablecomments{Magnitudes are measured using PSF-matched photometry with aperture correction.
Upper limits are based on the $2\sigma$ noise level in the corresponding aperture.
$z_{\rm photo}$ is from the type-I AGN model from \cite{Akins2024}.
}
\end{deluxetable*}

\begin{figure*}[ht!]
\epsscale{1.15}
\plotone{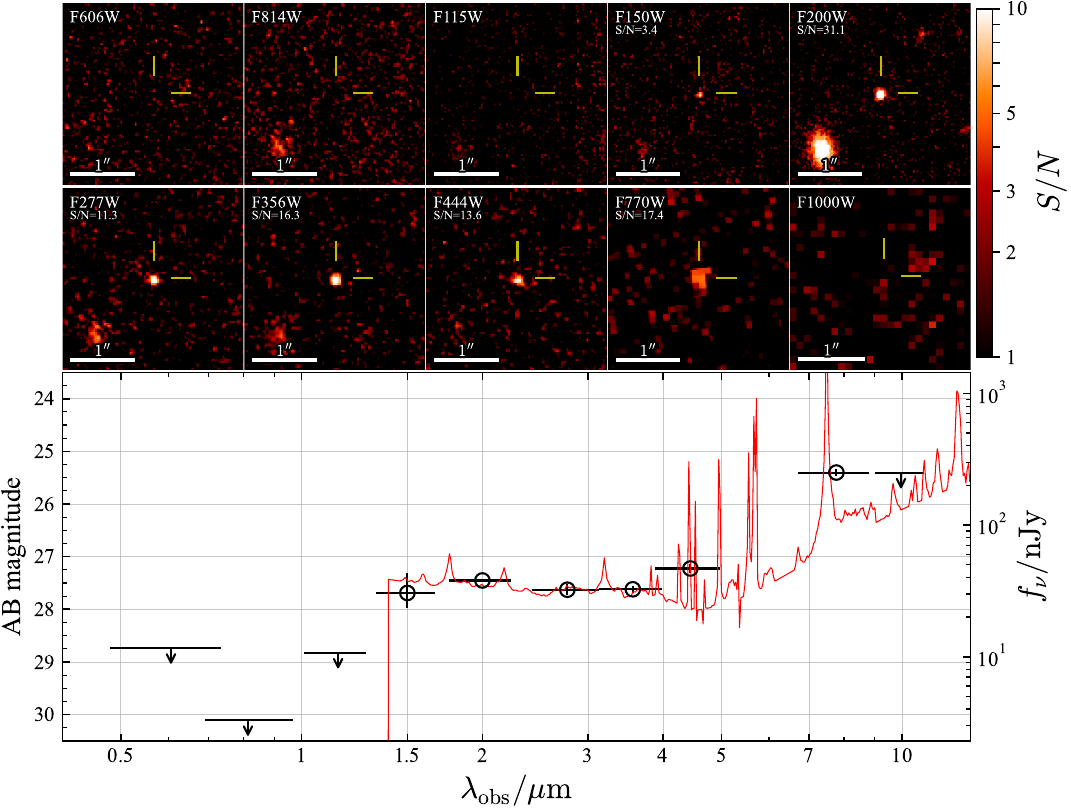}
\caption{
(Upper) HST + JWST $3^{\prime\prime} \times 3^{\prime\prime}$ cutout images centered on CW-LRD-z10.
Each image is scaled by a noise map.
If detected, the $S/N$ measured from forced photometry is shown in the top left corner.
(Lower) Observed SED of CW-LRD-z10.
For F606W, F814W, F115W, F1000W, and F2100W, we plot the $2\sigma$ upper limits.
The red line indicates the best-fit LRD model SED from the SED fitting analysis (Section\,\ref{ss:SED_fit})
\label{fig:sed_image}}
\end{figure*}

\section{Analysis of CW-LRD-z10}\label{s:analysis}

We describe the analysis of CW-LRD-z10, our single LRD candidate at $z\gtrsim10$.

\subsection{Forced photometry}\label{ss:forced_photo}

We perform PSF-matched forced photometry of CW-LRD-z10 from HST/F606W to MIRI/F770W, including filters not covered in COSMOS2025, to have uniform photometric measurements.
We reconstruct empirical PSFs based on field stars for each band (Section\,\ref{ss:image_fit}), and then match the PSFs of all images to the F770W PSF, the longest-wavelength filter (with the largest PSF FWHM) in which CW-LRD-z10 is detected.
We measure the photometry using a $0.\!\!^{\prime\prime}54$ diameter circular aperture, which corresponds to $2\times$ the FWHM of the F770W PSF.
The surface brightness profile of CW-LRD-z10 is well described by the PSF (Sections\,\ref{ss:image_fit} and \ref{ss:morph}); thus, we perform the aperture photometry with a correction to estimate the total photometry assuming the F770W PSF.
For non-detections in F606W, F814W, and F115W, we use error maps to compute the $2\sigma$ error on the aperture photometry as an upper limit. The forced photometry is consistent with the PSF photometry estimated from the image-based fitting analysis (Section\,\ref{ss:image_fit}) within the $1\sigma$ uncertainty.
For the non-detections in MIRI (F1000W and F2100W), we do not perform PSF matching to maintain a higher spatial resolution for the detection images.
To estimate upper limits in F1000W and F2100W, we use a $0.\!\!^{\prime\prime}66$ and $1.\!\!^{\prime\prime}35$ diameter circular aperture, respectively, which corresponds to $2\times$ each FWHM, and apply an aperture correction assuming each PSF.
The resulted photometry and upper limits are listed in Table.\,\ref{tab:info}.

\subsection{SED fitting}\label{ss:SED_fit}

Using forced photometry, we perform SED fitting to estimate the photometric redshift.
We employ the Bayesian SED modeling code \texttt{BAGPIPES} \citep{Carnall2018} for the following three setups:

\begin{itemize}
    \item a standard galaxy model having stellar mass of $\log\left(M_*/M_\odot\right)=4\,\mathchar`-\,12$, stellar metallicity of $Z_*/Z_\odot=0.1\,\mathchar`-\,1$, a delayed-$\tau$ star formation history with the timescale (\texttt{tau}) of $0.01\,\mathchar`-\,10\,{\rm Gyr}$ and \texttt{age} of $0.005\,\mathchar`-\,10\,{\rm Gyr}$, and dust attenuation ($A_V=0\,\mathchar`-\,8$).
    The upper limit on the \texttt{age} is automatically set to the age of the Universe at each redshift.
    We also include dust emission and nebular lines with $\log U=-2$.
    We fit the model in the redshift range of $z = 2\,\mathchar`-\,15$.
    This single high-$z$ galaxy model fitting is intended to confirm whether a normal galaxy SED can reproduce the observed SED.
    A poor fit would strengthen the interpretation that CW-LRD-z10 is an LRD, whose SED cannot be reproduced by a standard galaxy model alone, requiring a mixed model such as a dusty galaxy and a normal galaxy \citep[e.g.,][]{Labbe2023b} or scattered light \citep{Akins2024}.
    \item same galaxy model as above but limited to the redshift range of $z = 0\,\mathchar`-\,2$ to test the $z \sim 0.2$ solution.
    \item type-I AGN model from \cite{Akins2024} for fitting LRDs ($z = 0\,\mathchar`-\,15$), which incorporate a power-law accretion disk, flexible emission line templates based on \cite{Temple2021}, and dust attenuation with scattering.
\end{itemize}

By comparing the results with these three cases, we assess the possibility of interlopers or a non-LRD nature (SFGs or Balmer break galaxies).
We account for the non-detections in F606W, F814W, F115W, and F1000W by setting the flux to zero and the error to the $1\sigma$ upper limit.

Note that the type-I AGN model from \cite{Akins2024} includes hot torus emission at rest-frame MIR wavelengths, and the red slope is reproduced by invoking strong dust extinction. This may be inconsistent with recent studies reporting a lack of strong hot torus emission \citep{Wang2024a, Barro2024_comprehensive, Williams2024, Setton2025} and that the rest-frame optical slope of LRDs may be the Wien side of the black-body emission with $T_{\rm eff}\sim5000\,{\rm K}$ \citep{Inayoshi2025_binary, Kido2025, Liu_Hanpu2025, Lin2025_lowz}.
However, the model by \cite{Akins2024} can well reproduce the SED of LRDs from the rest-UV to rest-optical wavelengths.
Therefore, we exclude the F2100W photometry for SED fitting since it falls within the wavelength range where hot torus emission could have a significant impact when using the model by \cite{Akins2024}. Therefore, we only use the available HST, NIRCam, and MIRI (F770W, F1000W) photometry as inputs. Furthermore, we do not use any derived physical parameters other than the photometric redshift (e.g., $M_*$, SFR, $A_V$).

\subsection{Image-based fitting}\label{ss:image_fit}
To confirm CW-LRD-z10 as being compact, we conduct image modeling using \texttt{galight} \citep{Ding2020_HST}.
We fit a cutout image of $101\,{\rm pixel} \times 101\,{\rm pixel}$ centered on CW-LRD-z10 with either a point source or a Sérsic profile \citep{Sersic1963} with $r_e = 0.\!\!\,^{\prime\prime}03\,\mathchar`-\, 0.\!\!^{\prime\prime}5$ and $n = 0.5 \,\mathchar`-\,5.0$.

For PSF reconstruction, we follow the methodology described in \cite{Tanaka2024}.
First, we create an empirical PSF library for each filter based on field stars, which contains $\sim15\,\mathchar`-\,40$ single PSFs.
We test a fit of CW-LRD-z10 assuming each single PSF in the PSF library.  
Based on these single-PSF results, we sort the single PSFs in the order of $\chi^2$, then we stack the PSFs with the 5 lowest $\chi^2$ values and the 75\% lowest $\chi^2$ values.
We use these top-5 and the top-75\% stacked PSFs as the final PSFs.

To compare the fitting quality with PSF and Sérsic, we calculate the Bayesian Information Criterion (BIC, \citealt{Schwarz1978}):
\begin{equation}
    \mathrm{BIC} = -2\ln L + k \cdot \ln(N),
\end{equation}
where $L$ is the maximum likelihood, $k$ is the number of free parameters, and $N$ is the number of data points.
With the assumption of the Gaussian error, we calculate the term of $L$ as $-2\ln L = \chi^2$.
Generally, a lower BIC value indicates a better model fit, penalizing excessive model complexity.
We compute the difference in BIC as $\Delta \mathrm{BIC} = \mathrm{BIC}_{\rm PS} - \mathrm{BIC}_{\rm S\acute{e}rsic}$.  
If the Sérsic model yielded a BIC value lower than the PS model by more than 10 (i.e., $\Delta \mathrm{BIC} > 10$), we consider that the Sérsic model describes the data better than the point source model. Otherwise, the candidate is unresolved.

\section{Results}\label{s:results}

\begin{figure*}[ht!]
\epsscale{1.}
\plotone{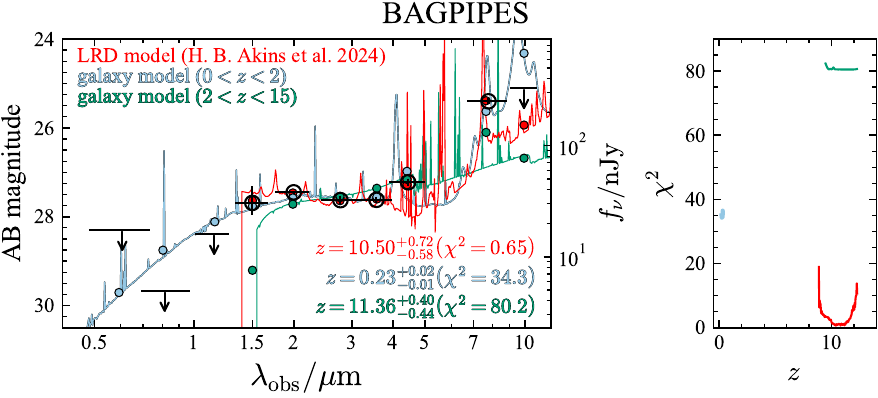}
\caption{
Observed photometry and SED fitting results using \texttt{BAGPIPES}.
In the left panel, black open circles, small dots, and lines represent the observed photometry, the model photometry, and the model SEDs.  
The estimated $z_{\rm photo}$ and corresponding $\chi^2$ value for each fit are indicated in the lower right corner.
The right panel shows the $\chi^2$ distribution as a function of redshift.
Red, cyan, and green lines indicate fits using the LRD model from \cite{Akins2024} over $0<z<15$, the galaxy model over $0<z<2$, and the galaxy model over $2<z<15$, respectively.
\label{fig:sedfit}}
\end{figure*}

\subsection{V-shape SED}\label{ss:V-shape_SED}
As shown in the Figure\,\ref{fig:sed_image} (lower), the photometry of CW-LRD-z10 indicates a V-shape SED: almost flat SED from F150W to F356W and a red steep slope from F356W to F770W.
This SED is consistent with the photometric redshift of $z\sim10$ (Section\,\ref{ss:photo-z}), because at this redshift the Balmer limit should shift to $4\,{\rm \mu m}$ and fall in the F444W coverage.
However, as discussed in \cite{Kocevski2024}, there are different interpretations of a red color ($m_{\rm F444W} - m_{\rm F770W} = 1.82\pm0.1$).
If we estimate the rest-optical slope from the $m_{\rm F444W} - m_{\rm F770W}$ color by simply assuming a power-law continuum, we obtain a value of $\beta = 1.05\pm0.18$.
Note that the photometry may include contributions from emission lines; thus, the derived slope above may not accurately reflect the intrinsic continuum slope.
In this section, we discuss possible non-LRD interpretations of the observed red $m_{\rm F444W} - m_{\rm F770W}$ color under different scenarios.

\subsubsection{Contribution of strong H$\alpha$}
One way to produce a large $m_{\rm F444W} - m_{\rm F770W}$ color without a red continuum would be an extremely strong H$\alpha$ line, which falls into the F770W coverage at $z\sim10$.
However, we rule out this scenario for the following reasons.
If we assume a flat continuum SED (constant $f_\nu$ at $f_{\rm F444W}$, i.e., $\beta=-2$) and attribute the entire F770W excess from F444W to H$\alpha$ emission, the inferred rest-frame H$\alpha$ EW is $\sim8000$\,{\AA}.
This EW is significantly larger than those predicted for metal-poor or metal-free galaxies ($\lesssim 4000$\,\AA; \citealt{Inoue2011}) or those of extreme emission line galaxies ($\sim 2000$\,\AA, e.g., \citealt{Withers2023}), making it difficult to explain within the framework of known galaxy populations.

For reference, JADES-GS-z14-0, discovered at $z_{\rm spec}=14.32$, was also detected in F770W, and the most reasonable interpretation of this bright F770W photometry is a contribution of strong H$\beta$+[O\,{\sc iii}]$\lambda\lambda4959, 5007$ lines \citep{Helton2025}.
However, JADES-GS-z14-0 exhibits a color of $m_{\rm F444W} - m_{\rm F770W} = 0.5$ \citep{Helton2025}, whereas our object shows a color of $m_{\rm F444W} - m_{\rm F770W} = 1.8$, suggesting that a different scenario is likely responsible.

If we assume an AGN-like continuum slope of $\beta=-1.5$, extremely strong H$\alpha$ with EW of $\sim 6000$\,{\AA} is still needed to reproduce the color, which significantly exceeds the predicted EWs for a seed BH ($\sim 1300$\,\AA, \citealt{Inayoshi2022}).
In contrast, assuming a steeper slope of $\beta = 0.5$ as typically observed in LRDs \citep[e.g.,][]{Greene2024, Lin2025_c3d} yields a more plausible EW of $800$\,{\AA}, consistent with the EWs of known LRDs \citep[e.g.,][]{Maiolino2024_chandra, Setton2024}.

\subsubsection{Balmer break}
If the Balmer break were extremely strong, it could produce a flux change around $4\,{\rm \mu m}$.
However, we can rule out the scenario in which a stellar population reproduces such a strong break, which matches the observed color.
Here, we simply model the SED of a Balmer break galaxy as a flat-$f_\nu$ spectrum with a sharp, wall-like break at a rest-frame wavelength of $0.3646\,{\rm \mu m}$, and quantify the strength of the break using the $f_\nu$ ratio across the break.
If the observed color of $m_{\rm F356W} - m_{\rm F770W} = 2.22$ in CW-LRD-z10 is fully described by the Balmer break, the implied Balmer break strength is $7.6$, which significantly exceeds the upper limit ($\sim3$) expected for normal stellar populations \citep[e.g.,][]{InayoshiMaiolino2024, deGraaff2025, Naidu2025}.

Such extremely strong Balmer breaks are found in some LRDs \citep{Williams2024, Labbe2024, Ji2025, Naidu2025, deGraaff2025}.
However, we can rule out this scenario for CW-LRD-z10 because it does not show $m_{\rm F277W}-m_{\rm F444W}$ and $m_{\rm F444W}-m_{\rm F770W}$ colors as large as the SED of The Cliff, as shown in Figure\,\ref{fig:cc}\,(right).
This is also evident in Figure\,\ref{fig:sed_image} (lower), which indicates that the observed SED of CW-LRD-z10 clearly does not match that of The Cliff \citep{deGraaff2025}.

Therefore, to explain the red color observed in CW-LRD-z10 across the $4\,\mathchar`-\,7\,{\rm \mu m}$, a rest-optical steep continuum slope as found in typical LRDs is required.
This conclusion is also consistent with the fact that the SED could not be well reproduced by a single galaxy model using \texttt{BAGPIPES} as described in Section\,\ref{ss:photo-z}.

\subsection{Compact morphology}\label{ss:morph}
As described in Section\,\ref{ss:image_fit}, we perform image-based morphological fitting across all filters from F150W to F770W where CW-LRD-z10 is detected.
In every band and for both the top-5 and the top-75\% PSFs, the point source model results in a lower BIC than the Sérsic profiles.
This result strongly supports the interpretation that the object is compact and unresolved at JWST resolution.
This compactness is further and independently confirmed in the flux-FWHM plane (Appendix\,\ref{ap:fwhm}), where CW-LRD-z10 follows the same sequence as the point source objects. The lower limit on the effective radius in our image-based fitting is $0.\!\!^{\prime\prime}03$, and we use this value as an upper limit for the effective radius.
The upper limit corresponds to a projected size of $\lesssim 120\,{\rm pc}$ at $z\sim10$, consistent with the size constraints of previous LRDs \citep[e.g.,][]{Furtak2023, Akins2024, Baggen2024}.

Recent high-redshift observations suggest that galaxies at $z\gtrsim10$ can be broadly divided into two populations: compact galaxies with strong N\,{\sc iv}] and spatially extended galaxies with weak N\,{\sc iv}] \citep{Harikane2025, Naidu2025_MoM}.
Our size upper limit is smaller than the typical rest-UV circularized half-light radius of the high-$z$ galaxy population ($\sim 200\,{\rm pc}$, \citealt{Ono2025}), and is instead consistent with the compact galaxy population ($\lesssim 100\,{\rm pc}$).
Therefore, at least, CW-LRD-z10 is morphologically distinct from the extended high-$z$ galaxy population.

One extreme hypothesis that might be considered here is that all of the high-$z$ compact population corresponds to LRDs. 
However, spectroscopically compact $z>10$ objects have distinct features from LRDs.
For example, GN-z11 and GHZ-2 do not show broad H$\alpha$ emission lines \citep{Alvarez2025_GNz11MIRI, Zavala2025}, MoM-z14 shows morphologies more extended than the PSF \citep{Naidu2025_MoM}, and GS-z14-1 does not show F770W excess compared to NIRCam photometry \citep{Wu2025_GSz141}.
Thus, this interpretation is unlikely.

If CW-LRD-z10 is a dwarf galaxy at $z\sim0.2$, the corresponding size limit would be $\lesssim100\,{\rm pc}$.
Typical dwarf galaxies with $M_* \sim 10^6M_\odot$ in the local Universe have effective radii of several hundred pc \citep[e.g.,][]{Norris2014}, which would be inconsistent with this upper limit on its size. 
Therefore, the size constraint disfavors typical extended dwarf galaxy solutions.
However, in the case of ultra-compact dwarf galaxies, the observed size constraint would remain consistent with $\lesssim100\,{\rm pc}$ \citep[e.g.,][]{Norris2014}.

\subsection{Photometric redshift}\label{ss:photo-z}
The SED of the selected object exhibits a distinct V-shape feature with a trough around F356W, consistent with the model SED of $z=10$ LRDs shown in Figure\,\ref{fig:depth}.
Additionally, the SED shows a flux discontinuity between F115W and F150W, consistent with the Lyman break falling between those filters, further supporting $z\sim10$.
While a typical dropout selection technique uses colors based on forced photometry such as $m_{\rm F115W}-m_{\rm F150W}$ or performs SED fitting in the selection stage, we apply only $S/N$ cuts to select F115W dropout objects (Section\,\ref{ss:color_selection}); thus, our dropout selection is looser compared to other dropout selections \citep[e.g.,][]{Harikane2023_highz, Casey2024, Franco2024}.
However, CW-LRD-z10 still passes the color criteria for F115W dropout galaxies proposed by \cite{Harikane2023_highz} (gray-filled region in Figure\,\ref{fig:cc} left).
Also, note that our selection does not rely solely on the F115W dropout, but also incorporates a requirement for a significant slope change across the Balmer limit around $\sim4\,{\rm \mu m}$, which is a distinctive feature of LRDs.
As illustrated by the green hexagon in Figure\,\ref{fig:cc} (right), a $z\sim2.5$ Balmer break galaxy, a typical interloper in F115W dropout selection, occupy a markedly different location from $z\sim10$ LRDs in the $m_{\rm F277W} - m_{\rm F444W}$ versus $m_{\rm F444W} - m_{\rm F770W}$ color-color plane, especially in $m_{\rm F444W} - m_{\rm F770W}$.
Thus, using both features makes our selection robust against $z\sim2.5$ Balmer break galaxies.

We test this $z\sim10$ LRD scenario through the SED fitting analysis described in Section\,\ref{ss:SED_fit}.
The $\chi^2$ distribution and the best-fit model SED are shown in Figure\,\ref{fig:sedfit}.
When fitting over the wide redshift range of $0<z<15$ with the type-I AGN model from \cite{Akins2024}, we obtain a best-fit photometric redshift of $z_{\rm photo}=10.50^{+0.72}_{-0.58}$.
Then, when fitting over the low redshift range of $0<z<2$ with the galaxy model, we obtain a best-fit photometric redshift of $z_{\rm photo}=0.23^{+0.02}_{-0.01}$.
The $z\sim10$ LRD solution has much lower $\chi^2$ with $\Delta \chi^2=33.7$ than the $z\sim0.2$ galaxy solution (see Figure\,\ref{fig:sedfit}).
The discrepancy between the observed photometry and the $z\sim0.2$ galaxy SED is particularly evident in F1000W, where the predicted flux density in the $z\sim0.2$ galaxy solution exceeds the observed upper limit due to the contribution from the PAH $7.6\,{\rm \mu m}$ feature.
In the $z\sim0.2$ galaxy case, the F115W dropout is attributed to the rest-optical-to-NIR dust-obscured continuum; however, this explanation fails, and it predicts brighter F814W and F115W photometry than the observed upper limits, resulting in further inconsistencies.
Thus, we conclude that the $z\sim0.2$ galaxy case can not reproduce the observed photometry for CW-LRD-z10.

Fitting over the high-redshift range of $2<z<15$ with the galaxy model yields photometric redshifts of $z_{\rm photo} = 11.36^{+0.40}_{-0.44}$ and higher $\chi^2$ values than the $z\sim10$ LRD solution ($\Delta \chi^2 = 79.6$).
\texttt{BAGPIPES} assumes a single galaxy population, and the resulting best-fit SED includes strong emission lines.
However, the high-redshift galaxy model fails to simultaneously reproduce both the rest-frame UV part and the rest-optical flux excesses in F444W and F770W.
This result suggests that the SED of CW-LRD-z10 is difficult to reconcile with a typical single-component galaxy model.
Therefore, we conclude that CW-LRD-z10 is highly likely to be an LRD at $z_{\rm photo}=10.50^{+0.72}_{-0.58}$, rather than a normal galaxy at $z \sim 0.2$ or $z\sim10$.

\subsection{UV luminosity}
If the photometric redshift is correct, CW-LRD-z10 would be one of the most distant SMBHs ever detected, comparable to GHZ-9 ($z_{\rm spec}=10.145\pm0.010$, \citealt{Kovacs2024_GHZ9, Napolitano2025_GHZ9}), UHZ-1 ($z_{\rm spec}=10.073\pm0.002$, \citealt{Bogdan2024_UHZ1, Goulding2023}), and GN-z11 ($z_{\rm spec}=10.6034\pm0.0013$, \citealt{Oesch2016, Jiang2021_GNz11, Bunker2023, Maiolino2023_GNz11}).
Assuming the photometric redshifts derived from the fits using the LRD model, we estimate the UV absolute magnitude using
\begin{equation}
M_{\rm UV} = m_{\rm 1450\left(1+z\right)} - 5\log_{10}\left(\frac{D_L}{10\,\mathrm{pc}}\right) + 2.5\log_{10}\left(1+z\right),
\end{equation}
where $D_L$ is the luminosity distance and $m_{\rm 1450\left(1+z\right)}$ is the rest-frame 1450\,{\AA} magnitude inferred from the best-fit model SED.
The resulting UV absolute magnitude is $M_{\rm UV} = -19.9^{+0.1}_{-0.2}$.
These results are further supported by the SED fitting performed with \texttt{EAZY} \citep{EAZY}, as detailed in Appendix\,\ref{ap:eazy}.
We also note that, given the lack of a definitive conclusion regarding the level of dust extinction in LRDs, we do not apply any correction for dust attenuation.
Therefore, the intrinsic UV absolute magnitude may be smaller than the value derived here.

\subsection{Non-detection in multi-wavelength data}
CW-LRD-z10 was also covered and not detected by ALMA Band 6 ($\sim250\,{\rm GHz}$) with a continuum sensitivity of $0.167\,{\rm mJy\,beam^{-1}}$ (CHAMPS ALMA program, ID: 2023.1.00180.L, PI: A. Faisst; A. Faisst et al. in prep.).
Based on the $z\sim0.2$ dusty dwarf SFG model SED, the expected continuum flux at ALMA Band 6 is $1.1\,{\rm \mu Jy}$, respectively, which falls significantly below the sensitivity limit. 
Therefore, this ALMA result does not rule out the $z\sim0.2$ solution.

We also check the Chandra/ACIS-I (COSMOS Chandra Merged Image Data Version 1.0, \citealt{Civano2016}) and VLA 3\,GHz (VLA-COSMOS, \citealt{Smolvic2017}), both downloaded from the COSMOS dataset in Infrared Processing and Analysis Center (IPAC, \citealt{COSMOS_IPAC}\footnote{\url{https://irsa.ipac.caltech.edu/data/COSMOS/overview.html}}).
CW-LRD-z10 is not detected in either image.
Chandra observations have the limiting depth of $2.2\times10^{-16}\,{\rm erg\,cm^{-2}\,s^{-1}}$ in the $0.5\,\mathchar`-\,2\,{\rm keV}$ band, and the VLA 3\,GHz observation has the 1$\sigma$ sensitivity of $2.3\,{\rm \mu Jy/beam}$.
These non-detections are consistent with previous studies that have reported the absence of X-rays \citep{Yue2024, Ananna2024, Maiolino2024_chandra, Akins2024, Lin2025_lowz, Sacchi2025} and radio \citep{Akins2024, Perger2024, mazzolari2024, Lin2025_lowz} emission in LRDs.

\subsection{Grism spectra}
As described in Section\,\ref{ss:grism}, CW-LRD-z10 has been observed by COSMOS-3D grism spectroscopy with NIRCam/F444W.
F444W ($3.86\,\mathchar`-\,4.99\,{\rm \mu m}$) can cover [O{\sc ii}]$\lambda\lambda$3726,3729, [Ne{\sc iii}]$\lambda$3869, [Ne{\sc iii}]$\lambda$3967, and H$\delta$ for a galaxy or LRD at $z\sim10$ and the PAH $3.3\,{\rm \mu m}$ for a $z\sim0.2$ galaxy.
However, CW-LRD-z10 with $m_{\rm F444W}=27.2$ is too faint to be detected in the COSMOS-3D spectrum, and indeed the resulting data show no significant line detections.
For comparison, \cite{Lin2025_c3d} reported line detections with line flux $S/N > 2$ only for sources with $m_{\rm F444W}<26$.

We conduct a mock observation based on the model SED shown in Figure\,\ref{fig:sedfit} with the JWST ETC\footnote{\url{https://jwst.etc.stsci.edu/}}.
The estimated $S/N$ is $<0.1$ even at the peak of [O\,{\sc ii}]$\lambda\lambda$3726,3729 in the best-fit LRD and galaxy model SED or the PAH $3.3\,{\rm \mu m}$ feature in the best-fit $z\sim0.2$ galaxy model.
Therefore, while the lack of emission-line detection is consistent with our SED fitting analysis, we cannot obtain any strong constraints on its redshift and nature from the current grism data.
Deeper follow-up spectroscopy will be essential to robustly confirm the redshift and the nature of this object.

\section{Discussion}\label{s:discussion}

\subsection{Luminosity function and evolution of the LRD Population} \label{ss:LF}

We loosely constrain the number density of the LRD at $z\sim10$ by assuming that CW-LRD-z10 is a genuine LRD at $z_{\rm photo}=10.5^{+0.7}_{-0.6}$ with $M_{\rm UV}=-19.9^{+0.1}_{-0.2}$ based on the results presented in the previous section.

We first evaluate the completeness of our selection, $C\left(M_{\rm UV}, z\right)$, as a function of $z$ and $M_{\rm UV}$ to estimate the luminosity function. 
We use the stacked model SED of LRDs from \cite{Akins2024} (shown in Figure\,\ref{fig:depth}) with a given $M_{\rm UV}$ and evolve the model SED from $z=5$ to $z=15$.
We then add photometric errors comparable to those in COSMOS-Web to these model SEDs to simulate mock observed photometry.
We check whether the mock photometry satisfies the selection criteria: $S/N < 2$ in F115W, $S/N > 5$ in F150W, F277W, F444W, and F770W, and color selections (equations\,\ref{eq:col1} - \ref{eq:col3}).
We evaluate completeness at three magnitude limits, $M_{\rm UV} = -21$, $-20$, and $-19$.
For each $(M_{\rm UV},z)$ combination, we perform Monte Carlo simulations with 10,000 samples to estimate $C\left(M_{\rm UV}, z\right)$.
Since the COSMOS-Web observations are mosaicked, the actual depth varies across the field (see \citealt{Casey2023, Franco2025_CW_NIRCam, Harish2025_CW_MIRI}).
In this study, we assume the depth distribution across the entire COSMOS-Web field and estimate the completeness for each depth accordingly.

\begin{figure}[ht!]
\epsscale{1.1}
\plotone{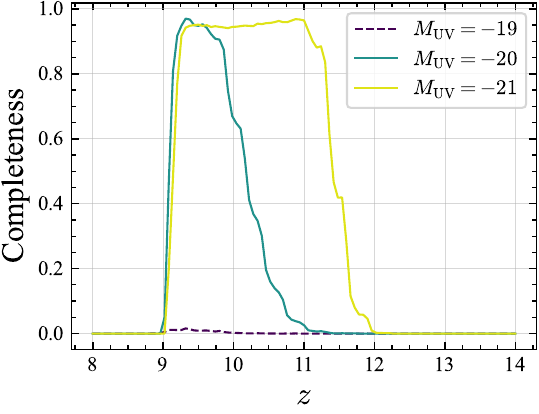}
\caption{
Completeness as a function of redshift and UV absolute magnitude, estimated from a Monte Carlo simulation of COSMOS-Web mock observations.
\label{fig:completeness}}
\end{figure}

The resulting completeness curves (only for the deepest regions) are shown in Figure\,\ref{fig:completeness}.  
Due to the limited survey depth, the completeness for $M_{\rm UV} = -19$ remains $\sim0$ across all redshifts, consistent with the our targeted luminosity range of $M_{\rm UV}\lesssim-19.5$.
Since our selection is based on the F115W-dropout criterion, the method becomes sensitive at $z\gtrsim9$ and achieves a completeness of $\sim 1$ around $z\sim10$ for sources with $M_{\rm UV} = -20$ and $-21$.  
For $M_{\rm UV} = -20$ ($-21$), the completeness declines rapidly above $z\sim10$ ($z\sim11$) as the LRDs become too faint to be detected in F150W.

\begin{figure*}[ht!]
\epsscale{1.15}
\plotone{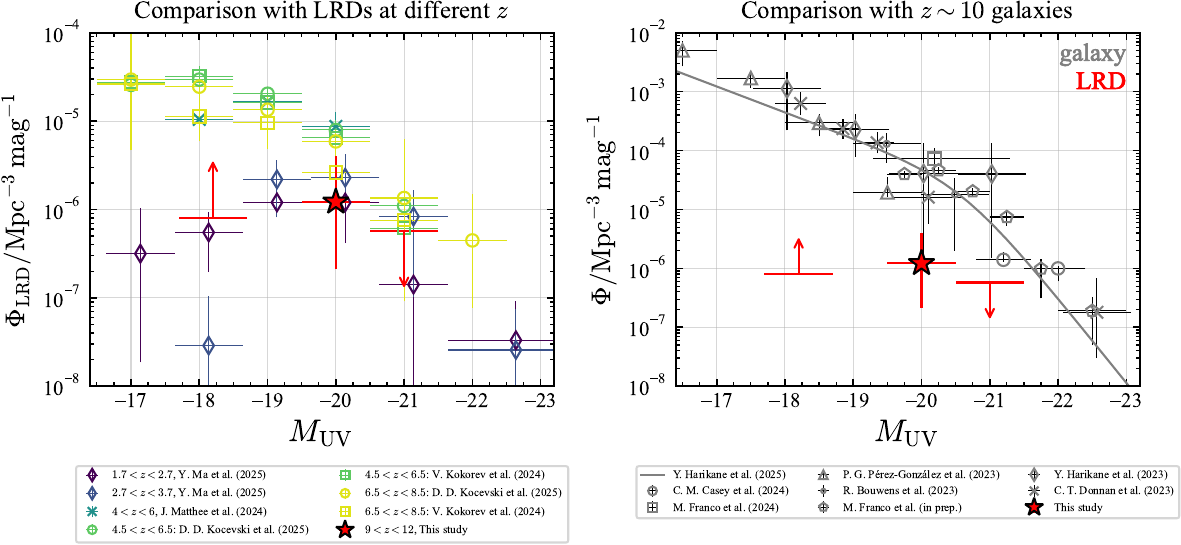}
\caption{
Luminosity function of $z\sim10$ LRDs shown by the red star data point.
(Left) Comparison with the luminosity functions of LRDs at lower redshifts ($1.7<z<8.5$) from previous studies \citep{Ma2025, Matthee2024, Kocevski2024, Kokorev2024_census}.
Note that the lower limit at $M_{\rm UV}=-18.2$ and $9<z<12$ is based on the assumption that only CAPERS-LRD-z9 \citep{Taylor2025} is a true $z\sim10$ LRD in the selection by \cite{Kocevski2024}, independently identified of the selection in this study.
(Right) Comparison of the LRD luminosity function with that of galaxies at $z\sim10$ (\citealt{Donnan2023, Harikane2023_highz, Franco2024, Casey2024, Bouwens2023, Perez2023_highz, Franco2025_dropout}).
Best-fit double power law model from \cite{Harikane2025} is indicated by the black line.
\label{fig:LF}}
\end{figure*}

\begin{figure*}[ht!]
\epsscale{1.}
\plotone{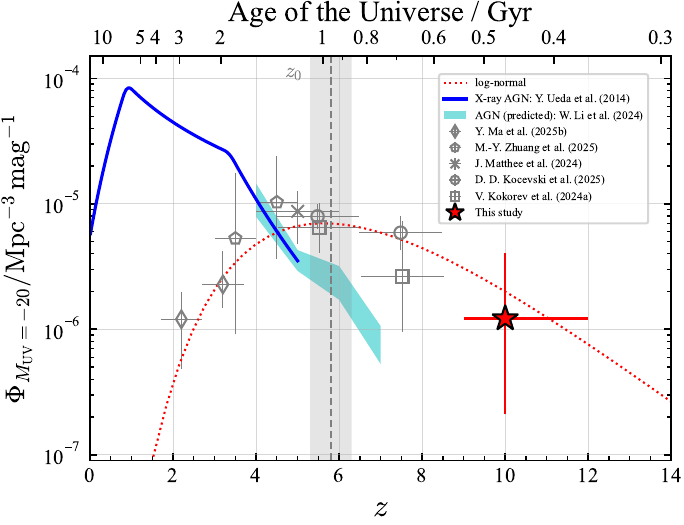}
\caption{
Redshift evolution of the luminosity function at $M_{\rm UV}=-20$.  
Symbols for lower-redshift LRDs \citep{Ma2025, Zhuang2025, Matthee2024, Kocevski2024, Kokorev2024_census} are the same as Figure\,\ref{fig:LF} (left).
The red dotted line indicates the best-fit log-normal distribution.
The gray dashed line and shaded region indicate the best-fit peak redshift ($z_0$).
Blue line and cyan shaded region indicate the AGN luminosity function at $M_{\rm UV}=-20$ from the X-rays \citep{Ueda2014} and model prediction \citep{Li2024_BHMF}.
Note that we assume an X-ray–UV luminosity ratio \citep{Lusso2016} and a photon index of $\Gamma=1.9$ to convert the X-ray \citep{Ueda2014} to UV luminosity function.
\label{fig:phi_z}}
\end{figure*}

We calculate the effective survey volume $V_{\rm eff}\left(M_{\rm UV}\right)$ for each $M_{\rm UV}$ as
\begin{equation}
    V_{\rm eff} \left(M_{\rm UV}\right) = \int C\left(M_{\rm UV}, z\right)\, \frac{{\rm d}V\left(z\right)}{{\rm d}z}\, {\rm d}z,
\end{equation}
where ${\rm d}V/{\rm d}z$ is the differential comoving volume and $C\left(M_{\rm UV}, z\right)$ is the completeness as evaluated above.
Based on this effective volume, we estimate the UV luminosity function of LRDs, $\Phi\left(M_{\rm UV}\right)$, using
\begin{equation}
    \Phi_{\rm LRD}\left(M_{\rm UV}\right) = \frac{N_{\rm LRD}\left(M_{\rm UV}\right)}{V_{\rm eff}\left(M_{\rm UV}\right)},
\end{equation}
where $N_{\rm LRD}\left(M_{\rm UV}\right)$ is the number of LRD candidates with $M_{\rm UV}$.
We estimate the luminosity function in three bins of $[-21.5,-20.5]$ and $[-20.5,-19.5]$. 
We evaluate the uncertainty based on Poisson statistics for the small number of objects \citep{Gehrels1986}. 
For the $M_{\rm UV} = -21$ bin where no candidates are identified, we adopt an upper limit at the $1\sigma$ ($84\%$) confidence level based on Poisson statistics \citep{Gehrels1986}.
The results are summarized in Table~\ref{tab:lf}.

Based on the effort in previous studies, we place constraints on the luminosity function at $M_{\rm UV} = -18.2$ for the same redshift range ($9<z<12$).
As introduced in Section\,\ref{s:intro}, CAPERS-LRD-z9 was photometrically selected in \citet{Kocevski2024} and spectroscopically confirmed in \citet{Taylor2025}.
\citet{Kocevski2024} derived the luminosity function using the $V_{\rm max}$ method \citep{Schmidt1968}.
Following this approach and assuming that only CAPERS-LRD-z9 was detected within the survey area by \cite{Kocevski2024} at $9 < z < 12$, the inferred number density would be $\Phi_{\rm LRD}(-18.2) > 8\times10^{-7}\,{\rm Mpc^{-3}\,{\rm mag^{-1}}}$.
Since \cite{Kocevski2024} detected other $z>10$ LRD candidates with similar luminosities ($M_{\rm UV} \sim -18$) that have not yet been spectroscopically confirmed, we consider this estimate as a lower limit on the low-luminosity side, and use it for comparison with our main results at $M_{\rm UV} < -19.5$.

\begin{deluxetable*}{lll}
\tablecaption{Estimated UV luminosity function of $z\sim10$ LRD ($z=9\,\mathchar`-\,12$) \label{tab:lf}}
\tablewidth{0pt}
\tablehead{
\colhead{$M_{\rm UV}$} & bin & \colhead{$\Phi_{\rm LRD}\left(M_{\rm UV}\right)/{\rm Mpc^{-3}\,mag^{-1}}$}
}
\startdata
$-20$ & $[-20.5,-19.5]$ & $\left(1.2^{+2.7}_{-1.0}\right)\times10^{-6}$ \\
$-21$ & $[-21.5,-20.5]$ & $<5.2\times10^{-7}$ \\
\enddata
\tablecomments{Uncertainty and upper limits are $1\sigma$ assuming Poisson statistics \citep{Gehrels1986}.}
\end{deluxetable*}

Figure\,\ref{fig:LF} (left) presents the UV luminosity function of LRDs at $z\sim10$ (red) with comparison to those at lower redshifts ($1.7<z<8.5$) from previous studies \citep{Ma2025, Matthee2024, Kocevski2024, Kokorev2024_census}.
\cite{Ma2025} provides the luminosity function with respect to the rest-frame $5500\,${\AA} magnitude ($M_{\rm 5500}$), thus we apply a correction of $M_{\rm UV} = M_{\rm 5500} + 1.86$ assuming the model SED from \cite{Akins2024}.
The aforementioned studies have found that the luminosity function of LRDs increases from $z\sim2$ to $z\sim6$, followed by a slight decline to $z\sim8$.
While we have only a single bin of $M_{\rm UV} = -20$ at $z=10$, the luminosity function at $M_{\rm UV}=-20$ is found to be lower than that at $z=6.5\,\mathchar`-\,8.5$ \citep{Kocevski2024} by $0.7^{+0.7}_{-0.5}$\,dex.

This redshift evolution is further illustrated in Figure~\ref{fig:phi_z}, where the co-moving space density at $M_{\rm UV} = -20$ is plotted as a function of $z$.
We fit the redshift evolution of the LRD luminosity function at $M_{\rm UV}=-20$ from our study and previous studies (\citealt{Ma2025}; \citealt{Zhuang2025}\footnote{For \cite{Zhuang2025}, we estimate the number density using their sample with $-19.5 < M_{\rm UV} < -20.5$.}; \citealt{Matthee2024}; \citealt{Kocevski2024}; \citealt{Kokorev2024_census}) with the log-normal distribution recently proposed by \citet{Inayoshi2025}, expressed as:
\begin{align}
    \Phi_{\rm LRD}\left(z\right) &= \Phi_{\rm 0,LRD} f\left(z\right) \nonumber\\
    &\hspace{20pt}\times\exp\left[-\frac{\left\{\ln\left(1+z\right)-\ln\left(1+z_0\right)\right\}^2}{2\sigma_z^2}\right], \label{eq:log-normal}\\
    f\left(z\right) &= \frac{\left(1+z\right)^{\frac{3}{2}}}{\left(s\left(1+z\right)^{\frac{1}{2}} - 1\right)^2},
\end{align}
where $\Phi_{\rm 0,LRD}$ is the normalization of the log-normal distribution, $z_0$ is the peak redshift, and $\sigma_z$ represents the width of the distribution.
The function $f(z)$ accounts for the redshift dependence of both the differential comoving volume element ${\rm d}V/{\rm d}z$ and the cosmic time interval ${\rm d}t/{\rm d}z$.
We adopt $s = 0.901$, corresponding to the assumed flat cosmology with $\Omega_m = 0.3$ and $\Omega_\Lambda = 0.7$.
The estimated parameters are $\Phi_{\rm 0,LRD}=\left(7.4^{+1.5}_{-1.5}\right)\times10^{-7}\,{\rm Mpc^{-3}\,mag^{-1}}$, $z_0=5.8^{+0.6}_{-0.5}$, and $\sigma_z=0.31^{+0.08}_{-0.06}$.
As shown in Figure~\ref{fig:phi_z}, the fitted log-normal distribution agrees well with the observed evolution, within the uncertainty.
The consistency with a log-normal distribution may support the scenario proposed in \cite{Inayoshi2025} in which the emergence of LRDs is closely linked to stochastic phenomena, for example, the initial accretion of seed BHs.
Such an interpretation, linking LRDs to seed BHs, is also consistent with the discovery of an LRD at $z \sim 10$, which indicates that LRDs were already present $\sim0.5\,{\rm Gyr}$ after the Big Bang.

While our log-normal fit is based only on the $M_{\rm UV} = -20$ luminosity bin, \cite{Inayoshi2025} fitted the same functional form to all LRDs with $M_{\rm UV} < -18$, obtaining a slightly higher peak redshift $z_0 = 6.53^{+0.04}_{-0.03}$ and a narrower width $\sigma_z = 0.218\pm0.005$.
As seen in Figure~\ref{fig:LF} (left), the luminosity functions at $z \lesssim 4$ from \cite{Ma2025} have peaks around $M_{\rm UV}\sim-20$ and concentrate more on the luminous side than those at $z\gtrsim4$.
Consequently, when fitting only the $M_{\rm UV} = -20$ bin, the relatively higher values at $z\lesssim4$ may make $z_0$ lower and $\sigma_z$ broader than the result in \cite{Inayoshi2025}.
This suggests that the shape of the luminosity function of LRDs is redshift-dependent, possibly indicating that either the physical conditions of accretion (i.e., Eddington ratio) in LRDs or the mass function of the LRD population evolves with redshift.

Recently, \cite{Hviding2025} suggested that while the contamination rate in photometrically selected LRDs is low ($\lesssim 10\%$), the completeness may be as low as $\sim50\%$ when compared to spectroscopically confirmed samples.
While we apply a completeness correction by calculating the effective survey volume in Section\,\ref{ss:LF}, it is important to note that this completeness estimate implicitly assumes the stacked SED of photometrically selected LRDs from \cite{Akins2024}.  
Therefore, the completeness with respect to the overall LRD population, particularly those with diverse or atypical SEDs, remains uncertain.
If the uncertainty in completeness is as large as the $\sim50\%$ as suggested by \cite{Hviding2025}, our inferred luminosity function could be underestimated by up to a factor of two.  
However, we emphasize that the uncertainty due to the small sample size dominates the overall uncertainty for this analysis.
Therefore, a larger sample with future surveys is a crucial next step to better constrain the evolutionary trend toward the early Universe.

\subsection{LRD fraction}\label{ss:LRD_fraction}
Figure~\ref{fig:LF}\,(right panel) presents the LRD luminosity function at $z \sim 10$ in comparison to the galaxy luminosity function at similar redshifts \citep{Donnan2023, Harikane2023_highz, Franco2024, Casey2024, Harikane2025}.
These allow us to estimate the LRD fraction of the galaxy population at $M_{\rm UV}=-20$ as,
\begin{equation}
    f_{{\rm LRD},M_{\rm UV}=-20}\left(z\right) = \frac{\Phi_{{\rm LRD},M_{\rm UV}=-20}\left(z\right)}{\Phi_{{\rm galaxy},M_{\rm UV}=-20}\left(z\right)}. \label{eq:lrd_frac}
\end{equation}
For the galaxy luminosity function, we use the fitted results from \cite{Harikane2022} for $z=3\,\mathchar`-\,6$ and \cite{Harikane2025} for $z>6$.
At $z\sim10$, we find that the LRD fraction is $2.6_{-2.1}^{+5.9}\%$.
In Figure~\ref{fig:LRD_fraction}, we show the redshift evolution of $f_{\rm LRD}$.
While $\Phi_{\rm LRD}$ peaks around $z\sim7$, the $f_{\rm LRD}$ increases monotonically from $z = 3$ to $z = 8$.
At $z=10$, despite a large uncertainty, there is no evidence of a sharp decline from $z=8$.
A linear fit with $f_{\rm LRD}\left(z\right) = a\log(1+z) + b$ yields $a=3.68^{+0.61}_{-0.60}$ and $b=-5.10^{+0.49}_{-0.50}$, suggesting that the LRD fraction could reach $f_{\rm LRD}\sim10\%$ at $z = 14$, where the current highest $z_{\rm spec}$ objects are found \citep{Carniani2024, Naidu2025_MoM}.

This increasing trend suggests that, although the number density of LRDs decreases toward higher redshift as described by the log-normal model (red dotted line in Figure\,\ref{fig:phi_z}), the number density of UV-bright galaxies declines even more rapidly than LRDs, causing the relative fraction of LRDs to increase.
We introduce two models, theoretical and empirical ones, to interpret this increasing trend quantitatively.

\subsubsection{Theoretical interpretation}
\begin{figure*}[ht!]
\epsscale{1.0}
\plotone{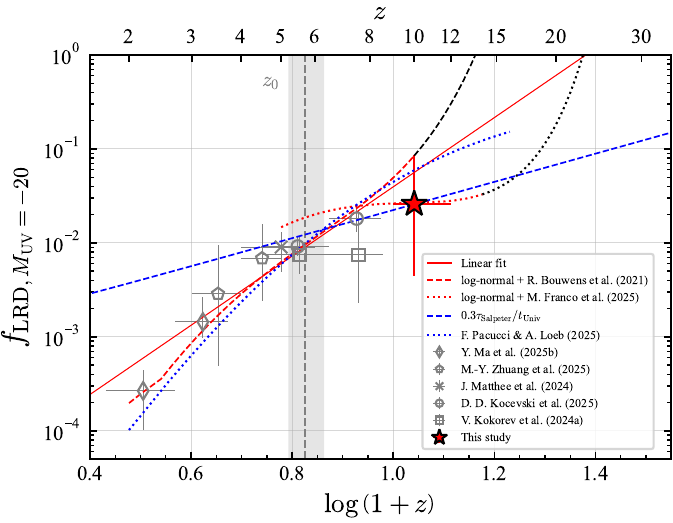}
\caption{
Redshift evolution of the LRD fraction relative to the overall galaxy population at $M_{\rm UV} = -20$.
Symbols for lower-redshift LRDs \citep{Ma2025, Zhuang2025, Matthee2024, Kocevski2024, Kokorev2024_census} are the same as Figure\,\ref{fig:LF} (left).
The red solid line shows the best-fit linear model.
The red dashed and dotted lines indicate the modeled $f_{\rm LRD}$ redshift evolution that combines the evolutions of the UV luminosity function of LRDs (the fitted log-normal distribution shown in Figure\,\ref{fig:phi_z}) and galaxies from \cite{Bouwens2021} and \cite{Franco2025_dropout}, respectively. 
For each model, the black line indicates the extrapolation of each luminosity function evolution beyond the range constrained by observational data. 
The blue dashed line indicates 0.3 times Salpeter time ($\tau_{\rm Salpeter}$, with the assumption of $\varepsilon=0.1$) divided by the age of the Universe ($t_{\rm Univ}$) at each redshift.
The blue dotted line represents the prediction from \cite{Pacucci2025}.
The gray dashed line and shaded region indicate the fitted peak redshift of the log-normal distribution ($z_0$).
\label{fig:LRD_fraction}}
\end{figure*}

In our theoretical model, we begin with simplified assumptions about the origins and evolution of LRDs and galaxies.
First, we assume that the formation rates of $M_{\rm UV}=-20$ LRDs ($\dot{n}_{\rm LRD}$) and galaxies ($\dot{n}_{\rm gal}$) are constant and independent of cosmic time.
Second, we assume that galaxies can maintain their luminosity after formation, while LRDs have a lifetime as LRDs, after which LRDs will not be observed as LRDs anymore.
Following the recently proposed BH envelope model by \cite{Kido2025}, we assume that an LRD lifetime corresponds to a Salpeter time \citep{Salpeter1964}, which is the $e$-folding timescale for $M_{\rm BH}$ evolution with the Eddington ratio of unity:
\begin{equation}
    \tau_{\rm Sal}\left(\varepsilon\right) \simeq 4.5\times10^7\left(\frac{\varepsilon}{0.1}\right)\,{\rm yr},
\end{equation}
where $\varepsilon$ is the radiative efficiency.
With these assumptions, the number of LRDs ($n_{\rm LRD}$) and galaxies ($n_{\rm gal}$) at cosmic age $t_{\rm Univ}$ evolve as:
\begin{align*}
    n_{\rm LRD}\left(t_{\rm Univ}\right) &= \tau_{\rm Sal}\left(\varepsilon\right) \times \dot{n}_{\rm LRD},\\
    n_{\rm gal}\left(t_{\rm Univ}\right) &= t_{\rm Univ} \times \dot{n}_{\rm gal}.
\end{align*}
Thus, the expected LRD fraction is:
\begin{equation}
    f_{\rm LRD}\left(t_{\rm Univ}\right) = \frac{n_{\rm LRD}\left(t_{\rm Univ}\right)}{n_{\rm gal}\left(t_{\rm Univ}\right)} = C \frac{\tau_{\rm sal}\left(\varepsilon=0.1\right)}{t_{\rm Univ}},
\end{equation}
where $C$ is a constant:
\begin{equation}
    C = \frac{\dot{n}_{\rm LRD}}{\dot{n}_{\rm gal}} \left(\frac{\varepsilon}{0.1}\right),
\end{equation}
including the formation rate ratio $\dot{n}_{\rm LRD}/\dot{n}_{\rm gal}$, uncertainty of the radiation efficiency $\varepsilon$, and possible deviation of the exact LRD lifetime from $\tau_{\rm Sal}$.
This model indicates that $f_{\rm LRD}$ evolves as the inverse of the cosmic age, i.e., $f_{\rm LRD} \propto t_{\rm Univ}^{-1}$.
Assuming $\varepsilon=0.1$ (i.e., $\tau_{\rm Sal}=4.5\times10^7\,{\rm yr}$), the model expectation with $C=0.3$ (blue dashed curve in Figure\,\ref{fig:LRD_fraction}) is broadly consistent with the the observed evolution of $f_{\rm LRD}$ at $z \gtrsim 5$.
However, at $z \lesssim 3$, the observed $f_{\rm LRD}$ significantly declines relative to the expectation.
One possible interpretation of this discrepancy is that the formation rate of LRDs decreases at $z \lesssim z_0$. 
$C=0.3$ may suggest that the formation rate ratio $\dot{n}_{\rm LRD}/\dot{n}_{\rm gal}$ is $0.3$ under the assumption of a radiation efficiency of $\varepsilon=0.1$.
However, LRDs may have higher radiation efficiencies of $\varepsilon \sim 0.3$, as suggested by \cite{InayoshiIchikawa2024}; therefore, no definitive conclusion can be drawn.

As an alternative theoretical framework, we also compared the results with the model of \cite{Pacucci2025}, in which LRDs are described as descendants of low-spin dark matter halos.
The predicted evolution of $f_{\rm LRD}$ at $M_{\rm UV}=-20$ (blue dotted curve in Figure\,\ref{fig:LRD_fraction}) agrees with previous observations within the uncertainties.
At $z=10$, the model prediction is higher than our $f_{\rm LRD}$ measurement, but still consistent within the uncertainty.
According to this model, $f_{\rm LRD}$ at $M_{\rm UV}=-20$ would reach $\sim0.1$ by $z=12$.
Again, our result has a large uncertainty, and it is still difficult to test and constrain the theoretical models strongly; therefore, we need to identify more LRD samples at $z\gtrsim10$.

\subsubsection{Empirical interpretation}
Following the definition of $f_{\rm LRD}$ (Equation~\ref{eq:lrd_frac}), we also model the redshift evolution of $f_{\rm LRD}$ by dividing the log-normal distribution of LRDs by the evolution of the number density of the galaxy population.
For the LRDs, we assume the log-normal distribution fitted to the observational results in Section~\ref{ss:LF} (the model shown in Figure\,\ref{fig:phi_z}).
For the galaxy population, we use the evolution of the high-$z$ UV luminosity function from \cite{Bouwens2021} and \cite{Franco2025_dropout}.
These works first model the galaxy UV luminosity function using the Schechter function \citep{Schechter1976}:
\begin{align}
    \Phi_{\rm galaxy} \left(M_{\rm UV}\right) &= \frac{\ln10}{2.5} \Phi^*_{\rm galaxy} 10^{-0.4\left(M_{\rm UV} - M_{\rm UV}^*\right)\left(\alpha+1\right)} \nonumber\\
    &\hspace{20pt}\times\exp\left[-10^{-0.4\left(M_{\rm UV} - M_{\rm UV}^*\right)}\right]
\end{align}
and provide the redshift evolution of the fitted parameters: $\Phi^*_{\rm galaxy}\left(z\right)$ (the normalization), $M^*_{\rm UV}\left(z\right)$ (the characteristic luminosity), $\alpha\left(z\right)$ (the faint-end slope).
Using these parameters, we derive the galaxy number density at $M_{\rm UV}=-20$ at each redshift, and compare it with the log-normal LRD distribution to investigate the redshift evolution of $f_{\rm LRD}$.

In \cite{Bouwens2021}, the evolution of the scaling ($\log\Phi^*_{\rm galaxy}$) of the galaxy luminosity function is modeled by a quadratic function of $z$, leading to a rapid decline in the number of galaxies at $z > 10$.
As a result, the model with \cite{Bouwens2021} predicts a sharp increase in the LRD fraction.
On the other hand, \cite{Franco2025_dropout} compiled JWST results and fitted $\log\Phi^*_{\rm galaxy}$ with a linear function of $z$.
This yields a less extreme evolution and more consistent results with observed trends than the model with \cite{Bouwens2021}.
Even under the model with \cite{Franco2025_dropout}, the LRD fraction does not decrease beyond $z_0$ (the peak redshift of the log-normal distribution and $=5.8$), providing a better match to the LRD fraction at $z\sim10$.
At $z \gtrsim 16$, the model with \cite{Franco2025_dropout} indicates that $M_{\rm UV}^*$ becomes larger than $-20$, resulting in a rapid decrease of the number density of galaxies at $M_{\rm UV} = -20$ and, consequently, a significant increase in the LRD fraction.
This increasing trend may suggest that SMBHs become luminous more rapidly than galaxies.
\cite{Inayoshi2025} hypothesized that LRDs represent the first or second accretion phase following seed black hole formation.
If this is the case, LRDs with $M_{\rm UV} = -20$ could emerge shortly after the seed BH formation.

As an extreme scenario, if we extrapolate the linear fit, log-normal + \cite{Bouwens2021}, and log-normal + \cite{Franco2025_dropout} models, each predicts that $f_{\rm LRD}>1$ at $z>24.0$, $13.5$, and $22.7$, respectively.
If this scenario proves to be correct, it would suggest that the first $M_{\rm UV}=-20$ objects in the Universe were LRDs.
Furthermore, these redshifts are consistent with theoretical predictions for the epoch of seed black hole formation.

These predictions are based on extrapolations from a log-normal distribution fitted to the number density derived from our candidate without spectroscopic confirmation, combined with the galaxy luminosity function; Thus, it is premature to draw firm conclusions given the significant uncertainties.
As this analysis is based on comparing the single luminosity bin of $M_{\rm UV}=-20$ and a single object, larger samples across a broader luminosity range will be necessary to robustly assess the LRD fraction across cosmic time.
Also, more importantly, it remains unclear whether the UV emission of LRDs originates from radiation associated with their SMBHs.
Therefore, the observed UV luminosity function may not directly trace the evolution of SMBHs.

\subsection{Bolometric luminosity of CW-LRD-z10}\label{ss:BHmass}
We estimate the bolometric luminosity ($L_{\rm bol}$) of CW-LRD-z10.
To infer $L_{\rm bol}$, we must make assumptions about the nature of the V-shaped SED and the SEDs in the MIR and longer wavelengths in which LRDs are typically undetected.
Initially, the red color observed in LRDs has been interpreted as originating from dust-obscured AGN or galaxies \citep[e.g.,][]{Akins2023, Barro2024, Labbe2023b}.
However, ALMA and JWST/MIRI observations have reported that LRDs lack hot torus emission and cold dust components, suggesting that LRDs would have low-level dust attenuation \citep[e.g.,][]{Setton2025, Casey2025_dust, Chen2025_dust}.
In this case, the V-shape SED is interpreted as the intrinsic SED of the LRDs without strong dust obscuration.
Additionally, strong Balmer breaks that a part of LRD show \citep{Williams2024, Labbe2024, Ji2025, Naidu2025, deGraaff2025} are not attributed to a stellar origin but to dense gaseous absorbers or envelopes with a high hydrogen column density of $\gtrsim$ a few $10^{23} {\rm cm}^{-2}$ \citep{InayoshiMaiolino2024, Ji2025}, supporting the interpretation that the red continuum is instead associated with radiation from an accreting BH.

In this study, we conservatively assume the lower limit of the $L_{\rm bol}$ of CW-LRD-z10 by assuming a dust-free case (no dust attenuation and re-emission).
We also adopt the interpretation suggested by recent studies that the blue excess originates from the host galaxy \citep[e.g.,][]{Naidu2025, Kido2025}; thus, we assume that the UV part does not contribute to the bolometric luminosity.
To estimate the intrinsic bolometric luminosity, we fit the red part of the SED (F356W, F444W, and F770W) with a blackbody spectrum assuming $T=5000\,\mathrm{K}$ and $A_V=0$ \citep{Kido2025, Inayoshi2025_binary, Liu_Hanpu2025, Lin2025_lowz}.
From the fitted results, the bolometric luminosity is estimated as $L_{\rm bol} = \left(2.9\pm 0.2\right)\times10^{44}\,\mathrm{erg\,s^{-1}}$.
With the $L_{\rm bol}$ and $\lambda_{\rm Edd}$, we estimate the $M_{\mathrm{BH}}$ as:
\begin{equation}
    \left(\frac{M_{\rm BH}}{M_\odot}\right) = \left(\frac{L_{\rm bol}}{1.26\times10^{38}\,{\rm erg\,s^{-1}}}\right) \left(\frac{\lambda_{\rm Edd}}{1.0}\right)^{-1}.
\end{equation}
While we do not know the exact $\lambda_{\rm Edd}$, we first assume an Eddington limit of $\lambda_{\rm Edd} < 1$ and obtain a lower bound $M_{\rm BH} \gtrsim 2\times10^6\,M_\odot$.
Thus, we may be witnessing an SMBH with mass slightly higher than that expected for the heavy seed scenario with direct-collapse BHs, $M_{\rm BH}\lesssim10^6M_\odot$.
The number density of $10^6M_\odot$ SMBHs at $z\sim10$ is predicted to be around $10^{-6}\,{\rm Mpc^{-3}\,dex^{-1}}$ in the BH mass function (\citealt{Li2024_BHMF}), which is roughly consistent with our UV luminosity function.
If, however, LRDs are accreting at super-Eddington rates as proposed in some studies \citep{Inayoshi2024, Lambrides2024, Pacucci2024_lrd, Madau2024_super_eddington, Liu_Hanpu2025}, the actual black hole mass could be significantly lower; the BH mass would decrease to $2\times10^5M_\odot$ with $\lambda_{\rm Edd}=10$.
Such BH masses of $\lesssim10^{6}M_\odot$ may be consistent with the scenario proposed by \citet{Inayoshi2025}, in which LRDs are in the phase shortly after the formation of heavy seed black holes, which are typically predicted to have masses of $10^{5\,\mathchar`-\,6}M_\odot$.

Nevertheless, this BH mass estimation carries substantial uncertainties, as it relies on various assumptions, including the shape of the SED and the Eddington ratio.
Since bright Balmer lines such as H$\alpha$ and H$\beta$ fall outside of the wavelength coverage of NIRSpec at $z\sim10$, deep spectroscopic follow-up with MIRI will be essential to obtain more robust constraints on $M_{\rm BH}$.

\subsection{Prospects}
Because NIRSpec only covers wavelengths up to the rest-frame $<0.5\,{\rm \mu m}$ at $z \sim 10$, deep spectroscopic follow-up observation with MIRI is essential for a robust confirmation of the redshift and physical nature of this candidate.
To reduce the uncertainty of the luminosity function (and LRD fraction) and enable the derivation of the luminosity function shape across different magnitude bins, we need to increase the sample size and expand the luminosity range of the sample.
One straightforward way to do this is to expand NIRCam-MIRI joint surveys in both area and depth.
F560W photometry can effectively help distinguish $z\sim10$ LRDs from the $z\sim0.2$ solution; therefore, increasing the number of MIRI filters in future surveys is helpful to robustly select LRD candidates.
Given the rarity of $z\gtrsim10$ galaxies, another efficient strategy to search for $z\gtrsim10$ LRDs is MIRI follow-up of $z\gtrsim10$ galaxy candidates found in NIRCam surveys.
To further improve the robustness of the selection and completeness estimates, it will also be important to construct a SED template that incorporates the diversity of LRD's SEDs, based on archival LRD spectra and upcoming spectroscopic data from ongoing spectroscopic campaigns such as EMBER (\href{https://www.stsci.edu/jwst/science-execution/program-information?id=7076}{GO\,7076}, PI: H. B. Akins).

Furthermore, a future survey project with the Galaxy Reionization EXplorer and PLanetary Universe Spectrometer (GREX-PLUS; \!\!\!\citealt{GREXPLUS_SB}; \citealt{GREXPLUS_SPIE2024}), a $1.0\,{\rm m}$ infrared space telescope for the 2030s, is expected to increase the sample size of bright high-$z$ LRDs.
The Deep survey of its wide-field camera with five filters covering $2\,\mathchar`-\,8\,{\rm \mu m}$ will cover an area of $10\,{\rm deg^2}$ with $5\sigma$ limiting magnitudes of $26.5\,\mathchar`-\,23.5$ mag \citep{GREXPLUS_SPIE2024}.
Combined with near-infrared deep survey data from Euclid or Roman, this survey will be capable of detecting LRDs with $M_{\rm UV} \lesssim -21.5$ and $\lesssim-22.5$ at $z \sim 10$ and $z \sim 13$, respectively, making it a valuable tool for exploring bright LRDs at $z\lesssim13$.
At $z > 15$, the wide-field camera only covers the rest frame $\lesssim0.5\,{\rm \mu m}$, making it difficult to distinguish LRDs from galaxies with strong emission lines or prominent Balmer breaks, similar to the limitations of NIRCam-only selections for $z\sim10$ LRDs (Section\,\ref{s:intro} and Appendix\,\ref{Ap:Kocevski}).
Therefore, complementary observations with MIRI covering $>8\,{\rm \mu m}$ will be crucial.
Bright galaxy candidates at $z \gtrsim 15$ identified by GREX-PLUS \citep{Harikane2022_HD, Harikane2025} should be followed up with MIRI, enabling the exploration of LRDs at even higher redshifts.

\section{Conclusion}\label{s:conclusion}
We conduct a joint color selection using both NIRCam and MIRI data to push the study of LRDs to $z\gtrsim10$.
The combination of NIRCam and MIRI enables the selection of LRD candidates at $z\sim10$, a redshift range that has been difficult to probe using previous NIRCam-only selection methods.
We demonstrate that the color-color diagram using F277W, F444W, and F770W (Figure~\ref{fig:cc} right) can efficiently separate $z\sim10$ LRDs from contaminants.
The key results are as follows:

\begin{itemize}
    \item By applying the selection with $S/N$ (an F115W dropout condition), colors (Equations\,\ref{eq:col1}, \ref{eq:col2}, and \ref{eq:col3}), and compactness, we identify one robust $z\sim10$ LRD candidate, CW-LRD-z10, from the $0.18\,{\rm deg^2}$ of the NIRCam-MIRI overlap region in COSMOS-Web.
    CW-LRD-z10 exhibits a clear V-shaped SED (Section\,\ref{ss:V-shape_SED}), an F115W dropout, and a compact morphology consistent with a point source (Section\,\ref{ss:morph}).
    \item A possible contaminant showing colors similar to those of $z \sim 10$ LRDs is a dusty SFG at $z \sim 0.2$, where PAH $3.3\,{\rm \mu m}$ and $5.6\,{\rm \mu m}$ features can boost the F444W and F770W fluxes.
    However, our SED fitting (Section\,\ref{ss:photo-z}) and morphological analysis (Section\,\ref{ss:morph}) strongly support the $z\sim10$ LRD interpretation.
    In particular, the non-detections in F814W, F115W, and F1000W are inconsistent with the $z\sim0.2$ dusty SFG case (Figure\,\ref{fig:sedfit}).
    \item The observed red color from F356W to F770W cannot be reproduced by $z \sim 10$ non-LRD galaxies with strong emission lines or stellar-origin Balmer breaks.
    To reproduce this color, not only a strong H$\alpha$ line but also a steep red continuum slope, characteristic of LRDs, is required (Section\,\ref{ss:V-shape_SED}).
    Therefore, we conclude that CW-LRD-z10 is highly likely to be a LRD at $z_{\rm photo}=10.50^{+0.72}_{-0.58}$ with the UV absolute magnitude of $M_{\rm UV}=-19.9^{+0.1}_{-0.2}$.
    \item Assuming CW-LRD-z10 is a genuine $z\sim10$ LRD, we derive the luminosity function of $z\sim10$ LRD at $M_{\rm UV} = -20$ in Section\,\ref{ss:LF} (Figure\,\ref{fig:LF}) as $\Phi_{\rm LRD}\left(M_{\rm UV}=-20\right) = \left(1.2^{+2.7}_{-1.0}\right)\times10^{-6}\,{\rm Mpc^{-3}\,mag^{-1}}$, consistent with a decline towards the highest redshifts from a peak at $z\sim6$ (Section\,\ref{ss:LF}, Figure\,\ref{fig:phi_z}).
    \item The fraction of LRDs among the overall galaxy population with $M_{\rm UV} = -20$ reaches $2.6^{+5.9}_{-2.1}\%$ at $z \sim 10$, intriguingly suggesting that an even higher LRD fraction may be expected at $z > 10$ (Section\,\ref{ss:LRD_fraction}).
    In an extreme scenario, extrapolating the evolution of the LRD fraction based on the number density evolution of both LRDs and galaxies indicates that the LRD fraction could approach $\sim1$ at $z \sim 20\,\mathchar`-\,30$ (Figure\,\ref{fig:LRD_fraction}).
    If this is the case, it may imply that SMBHs can become luminous more rapidly than galaxies, and that the very first $M_{\rm UV} = -20$ sources in the Universe may have been LRDs.
    \item Under the assumption of a blackbody SED with no dust attenuation and an Eddington-limited accretion, we estimate the BH mass to be $M_{\rm BH} \gtrsim 10^6 M_\odot$, placing this object among the most distant known active SMBHs and possibly witnessing the initial accretion phase of a seed black hole as suggested in \cite{Inayoshi2025} (Section\,\ref{ss:BHmass}).
\end{itemize}

Follow-up MIRI spectroscopy is essential to confirm the redshift, verify the nature of the source as an LRD, and better constrain the BH mass.
In the future, expanding MIRI surveys in terms of survey field and the number of filters, along with next-generation missions such as GREX-PLUS (\!\!\citealt{GREXPLUS_SB}; \citealt{GREXPLUS_SPIE2024}), which will explore much wider areas than JWST in the $2\,\mathchar`-\,8\,{\rm \mu m}$ range, will be crucial to increasing the sample of $z\gtrsim10$ LRDs.
These efforts will enable improved statistics and a more comprehensive sample over a broader luminosity range, providing key insights into the formation of SMBHs in the early Universe.

\begin{acknowledgments}
This work is based on observations made with the NASA/ESA/CSA James Webb Space Telescope.
The data were obtained from the Mikulski Archive for Space Telescopes at the Space Telescope Science Institute, which is operated by the Association of Universities for Research in Astronomy, Inc., under NASA contract NAS 5-03127 for JWST.
These observations are associated with program IDs 1727, 1287, 1837, 4233, 5398, 5893, and 6368, and the specific observations analyzed can be accessed via \dataset[DOI]{https://doi.org/10.17909/3fhv-y106}.
Some of the data products presented herein were retrieved from the Dawn JWST Archive (DJA). DJA is an initiative of the Cosmic Dawn Center (DAWN), which is funded by the Danish National Research Foundation under grant DNRF140.

Numerical computations were in part carried out on the iDark cluster, Kavli IPMU.
This work was made possible by utilising the CANDIDE cluster at the Institut d’Astrophysique de Paris. The cluster was funded through grants from the PNCG, CNES, DIM-ACAV, the Euclid Consortium, and the Danish National Research Foundation Cosmic Dawn Center (DNRF140). It is maintained by Stephane Rouberol.
We thank Anna de Graaff for guiding how to access the data and Kunihito Ioka, Pablo G. Pérez González, and Fabio Pacucci for constructive discussions.
We thank the anonymous referee for helpful feedback.
Kavli IPMU is supported by World Premier International Research Center Initiative (WPI), MEXT, Japan.
The Cosmic Dawn Center (DAWN) is funded by the Danish National Research Foundation under grant DNRF140.
TST is supported by Japan Society for the Promotion of Science (JSPS) KAKENHI Grant Number JP25KJ0750, a grant from the Hayakawa Satio Fund awarded by the Astronomical Society of Japan, and the Forefront Physics and Mathematics Program to Drive Transformation (FoPM), a World-leading Innovative Graduate Study (WINGS) Program at the University of Tokyo.
YH is supported by JSPS KAKENHI Grant Number JP24H00245 and JP22K21349.
KI acknowledges support from the National Natural Science Foundation of China (12233001), the National Key R\&D Program of China (2022YFF0503401), and the China Manned Space Program (CMS-CSST-2025-A09).
JTS is supported by the Deutsche Forschungsgemeinschaft (DFG, German Research Foundation) - Project number 518006966.
MO is supported by the JSPS KAKENHI Grant Number JP24K22894.
MA is supported by FONDECYT grant number 1252054, and gratefully acknowledges support from ANID Basal Project FB210003 and ANID MILENIO NCN2024\_112.
ET acknowledges support from ANID Fondecyt Regular grants 1241005 and 1250821, and CATA BASAL program FB210003.
SEIB is supported by the Deutsche Forschungsgemeinschaft (DFG) under Emmy Noether grant number BO 5771/1-1.
MF is supported by the European Union's Horizon 2020 research and innovation programme under the Marie Skłodowska-Curie grant agreement No 101148925.
DBS gratefully acknowledges support from NSF Grant 2407752.
\end{acknowledgments}

\appendix
\section{High-redshift LRDs in D. D. Kocevski et al. (2024)}\label{Ap:Kocevski}
\cite{Kocevski2024} selected LRD candidates at $2<z_{\rm photo}<11$ based on the slope of the SED using only NIRCam photometry.
Among them, 13 objects listed in Table~\ref{tab:kocevski} has the best-fit photometric redshift of $z_{\rm photo}\ge10$ in \cite{Kocevski2024}.
We examine archival JWST spectroscopic data and find that archival spectra already exist for 6 out of the 13 candidates from \href{https://www.stsci.edu/jwst/science-execution/program-information?id=1287}{GTO\,1287} (PI: K. Isaak), \href{https://www.stsci.edu/jwst/science-execution/program-information?id=4233}{GO\,4233} (RUBIES, PI: A. de Graaff and G. Brammer), and \href{https://www.stsci.edu/jwst/science-execution/program-information?id=6368}{GO\,6368} (CAPERS, PI: M. Dickinson).
While we do not provide detailed analysis for each object here as it is beyond the scope of this paper, we perform a quick look at processed data in the Mikulski Archive for Space Telescopes (MAST\footnote{\url{https://mast.stsci.edu/portal/Mashup/Clients/Mast/Portal.html}}) and the DJA, and manually estimate $z_{\rm spec}$.
The confirmed $z_{\rm spec}$ is summarized in Table\,\ref{tab:kocevski}.
Three of them have $z_{\rm spec}<9$ ($5.93$, $6.07$, and $8.50$), and the three are at $9<z_{\rm spec}<10$ ($9.29$, $9.65$, and $9.94$).
Thus, simply using these numbers, the success rate for a genuine $z_{\rm spec}\sim10$ object in the $z_{\rm photo}>10$ sample in \cite{Kocevski2024} is $50\%$.

Among the three $9<z_{\rm spec}<10$ objects, one is CAPERS-LRD-z9 spectroscopically confirmed as an LRD in \cite{Taylor2025}.
Then, one is CEERS-23931 at $z_{\rm spec}=9.94$ already reported as CAPERS-EGS-25297 in \cite{Pollock2025} and \cite{Donnan2025}.
CEERS-23931 has an optical continuum slope of $\beta=-0.65$ ($f_\lambda\propto\lambda^\beta$, \citealt{Donnan2025}), not spectroscopically showing a sign of the steep red continuum.
Instead, this object has extremely strong emission lines, such as [O\,{\sc ii}]$\lambda\lambda$3726,3729, [Ne\,{\sc iii}]$\lambda$3869, [Ne\,{\sc iii}]$\lambda$3967+$H\epsilon$, H$\delta$, and H$\gamma$+[O\,{\sc iii}]$\lambda$4363, and these strong emission lines boosts the F410M and F444W photometry and mimic the V-shape color in NIRCam photometry (see \citealt{Donnan2025}).

\begin{figure}[ht!]
\epsscale{1.13}
\plotone{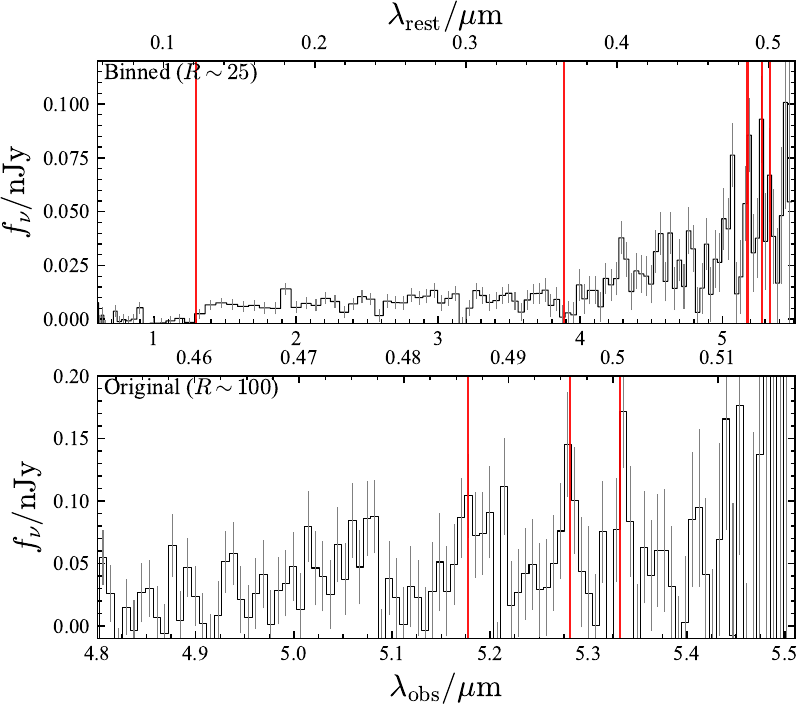}
\caption{
Spectrum of JADES-67592 ($z_{\rm spec}=9.65$) downloaded from DJA.
The upper and lower panels show the overall four times binned PRISM spectra ($R\sim25$) and the zoom-in original ($R\sim100$) spectra around H$\beta+$[O\,{\sc iii}]$\lambda\lambda$4959, 5007.
The wavelengths of Ly$\alpha$, Balmer limit, H$\beta$, and [O\,{\sc iii}]$\lambda\lambda$4959, 5007 are shown in red vertical lines.
\label{fig:Kocevski_spec}}
\end{figure}

The other object, JADES-67592 at $z_{\rm spec}=9.65$ with $M_{\rm UV} = -17.12$.
Due to the faintness, it is challenging to confirm the presence of a broad H$\beta$ line (Figure\,\ref{fig:Kocevski_spec}).
JADES-67592 spectroscopically exhibits a Balmer break as shown in Figure\,\ref{fig:Kocevski_spec}, and the strength of the Balmer break, defined as a $f_\nu$ ratio between the rest-frame wavelength ranges of $3620\,\mathchar`-\,3720\,{\rm \AA}$ and $4000\,\mathchar`-\,4100\,{\rm \AA}$, is $\sim 2$.
This break strength can be reproduced by a stellar-origin Balmer break.
JADES-67592 is not detected in either F560W or F770W from SMILES (\href{https://www.stsci.edu/jwst/science-execution/program-information?id=1207}{GTO\,1207}, PI: G. Rieke, \citealt{Rieke2024}) due to its faint nature ($M_{\rm UV}=-17.5$).
Given the 5$\sigma$ detection limits for a point source of SMILES, $f_{\rm F560W}=0.21\,{\rm \mu Jy}$ and $f_{\rm F770W}=0.20\,{\rm \mu Jy}$ \citep{Alberts2024}, the color upper limit of $m_{\rm F444W} - m_{\rm F770W} < 3$ is not stringent enough to robustly distinguish between a Balmer break galaxy and an LRD at $z\sim10$ (see Figure\,\ref{fig:cc}).
Thus, with the currently available data, we can not confirm if JADES-67592 is a genuine LRD or not.
To confirm the nature, deep MIRI photometry is needed to distinguish between an LRD and a galaxy with a stellar-origin Balmer break.
It is also important to obtain a deeper NIRSpec spectrum to confirm the presence of a broad H$\beta$ line.

These results indicate that, without incorporating MIRI F770W data, there is a risk of selecting contaminant sources that are not genuine $z\sim10$ LRDs, highlighting the advantage of our NIRCam-MIRI joint selection method in reliably identifying $z\sim10$ LRDs.
Also note that all of the spectroscopically unconfirmed $z_{\rm photo}>10$ LRD candidates in \cite{Kocevski2024} have $M_{\rm UV} \gtrsim -19$.
Thus, our candidate, CW-LRD-z10, with $M_{\rm UV} = -19.9$, discovered in a relatively shallow but wide-area field, complements their candidates by probing the brighter end of the LRD population.

\begin{deluxetable*}{llllllll}
\tablecaption{$z_{\rm photo}>10$ candidates reported in \cite{Kocevski2024}\label{tab:kocevski}}
\tablewidth{0pt}
\tablehead{
\colhead{ID} & \colhead{R.A. [deg]} & \colhead{Decl. [deg]} & \colhead{$M_{\rm UV}$} & \colhead{$z_{\rm photo}$} & \colhead{$z_{\rm spec}$} & \colhead{PID} & \colhead{Comments}
}
\startdata
CEERS-23931       & 214.817112 & $52.748343$ & $-19.77$ & 10.21 & 9.94 & 6368 & Non-LRD, EGS-25297 \citep{Donnan2025} \\
JADES-67592       & 53.072478  & $-27.855352$ & $-17.12$ & 11.26 & 9.65 & 1287 & Not detected in F560W and F770W \\
JADES-82209       & 53.197512  & $-27.782456$ & $-18.05$ & 10.48 & - & & Detected in F115W \\
PRIMER-COS-46911  & 150.175418 & $2.234439$  & $-18.84$ & 10.60 & - & &  \\
PRIMER-COS-70533  & 150.136254 & $2.30803$   & $-18.11$ & 11.92 & 9.29 & 6368 & CAPERS-LRD-z9 \citep{Taylor2025}\\
PRIMER-COS-78729  & 150.161027 & $2.465804$  & $-19.02$ & 10.45 & 5.93 & 6368 & Detected in F090W and F115W \\
PRIMER-COS-113415 & 150.155705 & $2.33761$   & $-17.58$ & 10.69 & - & & Not detected in F770W \\
PRIMER-UDS-66751  & 34.254151  & $-5.290015$ & $-18.28$ & 10.54 & 6.07 & 4233 & tentatively detected in F115W \\
PRIMER-UDS-106389 & 34.426489  & $-5.225494$ & $-19.29$ & 11.29 & - & & Not covered with F770W \\
PRIMER-UDS-134793 & 34.505257  & $-5.176794$ & $-18.01$ & 10.57 & - & & Not covered with F770W \\
PRIMER-UDS-151408 & 34.404447  & $-5.104179$ & $-18.91$ & 11.38 & - & & Not covered with F770W \\
PRIMER-UDS-151980 & 34.422908  & $-5.104757$ & $-18.80$ & 10.48 & - & & Not detected in F770W \\
PRIMER-UDS-173887 & 34.238463  & $-5.140396$ & $-18.81$ & 10.00 & 8.50 & 4233 & Detected in F770W \\
\enddata
\tablecomments{
ID, R.A., Decl., $M_{\rm UV}$, and $z_{\rm photo}$ are from Table\,3 in \cite{Kocevski2024}.
Note that, for objects with available $z_{\rm spec}$, we update $M_{\rm UV}$ with the value calculated with $m_{\rm UV}$, $\beta_{\rm UV}$, $z_{\rm photo}$, and $z_{\rm spec}$. 
}
\end{deluxetable*}

\section{Objects excluded in the visual insepction}\label{ap:visual_inspection}
\begin{figure*}[ht!]
\epsscale{1.16}
\plotone{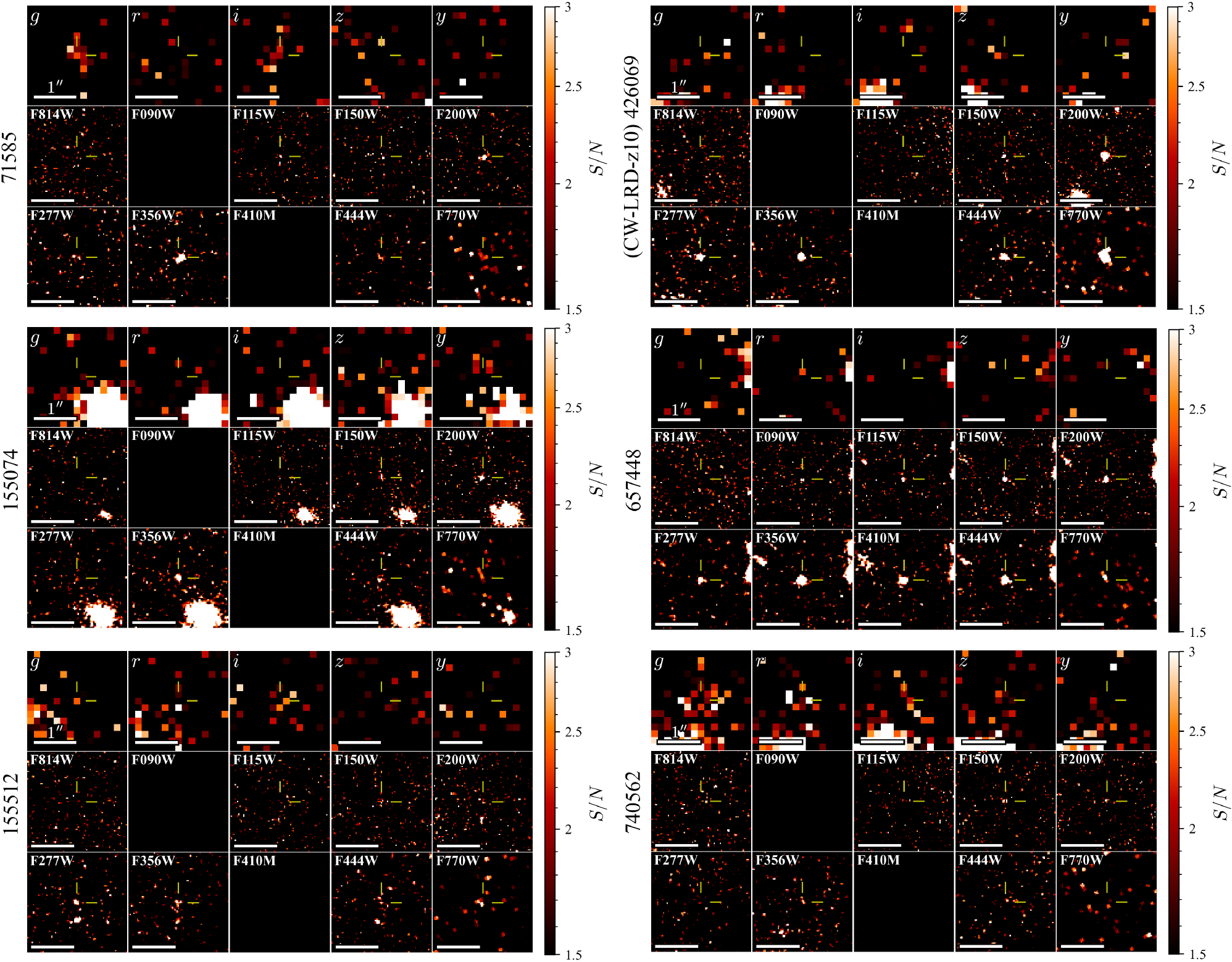}
\caption{
HSC ($grizy$), HST (F814W), and JWST $2.\!\!^{\prime\prime}5 \times 2.\!\!^{\prime\prime}5$ cutout images centered on each object excluded in visual inspection.
For comparison, we also plot CW-LRD-z10 (\texttt{ID:426069}).
Each image is scaled by a noise map.
\label{fig:vi_excluded}}
\end{figure*}

As described in Section\,\ref{ss:color_selection} and summarized in Table\,\ref{tab:numbers}, five objects that pass the selection criteria based on $S/N$, color, and compactness are excluded through visual inspection, and only CW-LRD-z10 is selected as a final candidate.
In this appendix, we show images of the five excluded objects listed in Table\,\ref{tab:vi} to demonstrate why they are unlikely to be $z\sim10$ LRDs.
Figure\,\ref{fig:vi_excluded} shows cutout images of these sources (and CW-LRD-z10 for comparison) normalized by the noise map.
These objects are generally excluded because they are not clearly detected in the F770W band.
Their F770W images are indistinguishable from the surrounding noise, suggesting that their F770W photometry may have been strongly influenced by noise.
Indeed, in COSMOS2025, the F770W signal-to-noise ratios ($S/N_{\rm F770W}$) for all objects except CW-LRD-z10 (\texttt{ID:426069}) are $\lesssim10$, significantly lower $S/N$ values compared to CW-LRD-z10.

Also note that some galaxy candidates are detected in HSC or F115W when combining all available JWST imaging data (i.e., not limited to COSMOS-Web alone).
\texttt{ID:71585} is detected in HSC and F115W, and \texttt{ID:657448} is detected in F090W and F115W from the combined PRIMER and COSMOS-Web data, suggesting that these are likely low-redshift interlopers.
If all four remaining candidates are $z \sim 10$ LRDs, the number density with $M_{\rm UV}=-20$ estimated in Section\,\ref{ss:LF} would increase to $\Phi_{\rm LRD} = \left(4.7^{+3.7}_{-2.2}\right)\times10^{-6}\,{\rm Mpc^{-3}\,{\rm mag^{-1}}}$.
Correspondingly, the LRD fraction at $z \sim 10$ calculated in Section\,\ref{ss:LRD_fraction} would increase to $f_{{\rm LRD},M_{\rm UV}=-20} = 9.9^{+7.8}_{-4.7}\%$.
However, given that their appearance is indistinguishable from noise in the F770W image, it is unlikely that these F770W detections are real.
Future spectroscopic confirmation will be essential to verify whether these candidates are indeed not $z \sim 10$ LRDs.

\begin{deluxetable*}{lllllll}
\tablecaption{Objects excluded in visual inspection process\label{tab:vi}}
\tablewidth{0pt}
\tablehead{
\colhead{ID} & \colhead{R.A. [deg]} & \colhead{Decl. [deg]} & \colhead{$z_{\rm photo}$} & $S/N_{\rm F770W}$& $M_{\rm UV}$& \colhead{Comments}
}
\startdata
71585 & 149.9223103 & $2.166595859$ & $0.23^{+0.10}_{-0.08}$ & 11.0 & $-19.8$ & Detected in HSC $g$\\
155074 & 150.2131795 & $2.133412186$ & $9.3^{+0.7}_{-1.1}$ & 9.8 & $-20.3$ & Not clearly detected in F770W\\
155512 & 150.2246714 & $2.137034303$ & $0.12^{+0.05}_{-0.07}$ & 12.4 & $-19.5$ & Not clearly detected in F150W, F444W, and F770W \\
426069 & 150.0841004 & $2.453894622$ & $0.17^{+0.01}_{-0.01}$ & 31.6 & $-19.9$ & Reported as CW-LRD-z10 \\
657448 & 150.0995828 & $2.337128099$ & $8.5^{+0.3}_{-0.7}$ & 7.9 & $-19.2$ & Detected in F115W (PRIMER) \\
740562 & 150.2236736 & $2.223625423$ & $0.2^{+2.0}_{-0.2}$ & 6.0 & $-20.3$ & Not clearly detected in F770W \\
\enddata
\tablecomments{
ID, R.A., Decl., $z_{\rm photo}$, and $S/N_{\rm F770W}$ are from COSMOS2025 catalog \citep{Shuntov2025_COSMOS2025}.
$M_{\rm UV}$ is calculated from F150W photometry and the assumption of $z=10$.
}
\end{deluxetable*}

\newpage 
\section{Flux-FWHM relationship suggesting a compact morphology}\label{ap:fwhm}
\begin{figure}[ht!]
\epsscale{1.1}
\plotone{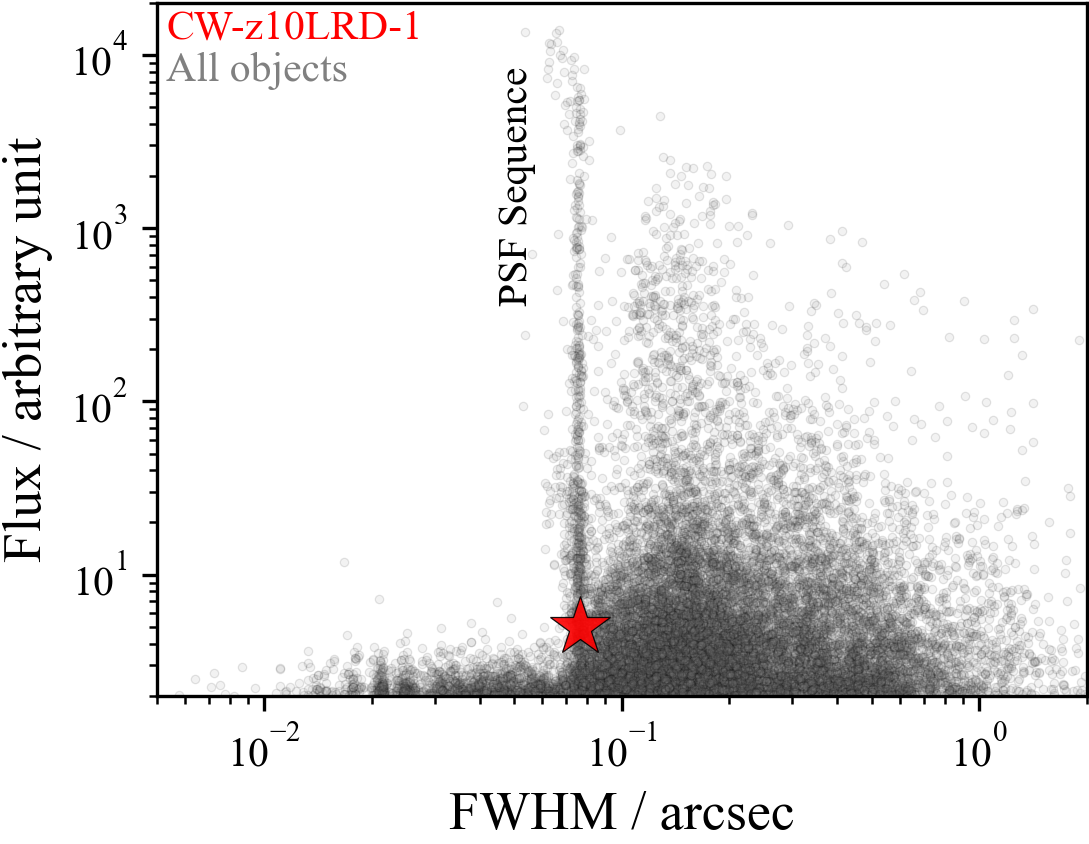}
\caption{
Flux as a function of FWHM for all detected objects in a single tile of the F200W image ($\sim 0.019\,{\rm deg^2}$), measured using \texttt{measure\_FWHM} function in \texttt{galight}.  
CW-LRD-z10 exhibits an FWHM consistent with the sequence of point sources, strongly suggesting that it is a very compact object similar to previously found LRDs.
\label{fig:size_flux}}
\end{figure}

To confirm the compact morphology independently of the image-based fitting (Section\,\ref{ss:image_fit}), we examine the relationship between flux and FWHM for objects within the F200W image from COSMOS-3D, where CW-LRD-z10 is detected with the highest signal-to-noise ratio among all filters due to its relatively long exposure time.
The flux is approximately measured using \texttt{DAOStarFinder} in \texttt{photutils}, witch is based on based on the DAOFIND algorithm \citep{Stetson1987}.
We set the FWHM of a 2D Gaussian kernel (\texttt{fwhm} parameter in \texttt{DAOStarFinder}) to 4 pixels.
For each sources detected by \texttt{DAOStarFinder}, we measure FWHM using the \texttt{measure\_FWHM} function in \texttt{galight}.
Figure\,\ref{fig:size_flux} shows that CW-LRD-z10 exhibits an FWHM consistent with the sequence of objects following point sources.
This result supports that a point source well describes the morphology of CW-LRD-z10, as suggested from the image-based morphological fitting (Section\,\ref{ss:morph}).

\section{SED fitting with \texttt{EAZY}}\label{ap:eazy}
\begin{figure*}[ht!]
\epsscale{1.}
\plotone{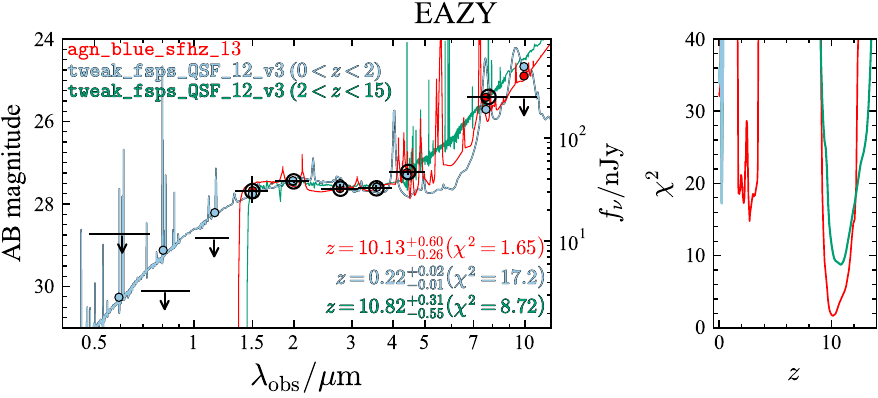}
\caption{
Same as Figure\,\ref{fig:sedfit} but for the results with \texttt{EAZY}.  
Red, cyan, and green correspond to fits using the \texttt{sorted\_agn\_blue\_sfhz\_13} template over $0<z<15$, the \texttt{tweak\_fsps\_QSF\_12\_v3} template over $0<z<2$, and the \texttt{tweak\_fsps\_QSF\_12\_v3} template over $2<z<15$, respectively.
\label{fig:sedfit_eazy}}
\end{figure*}

Generally, the results of SED fitting analyses depend on the choice of templates or models \citep[e.g.,][]{Wuyts2009, Newman2025}.
To minimize dependencies on a specific method, we also utilize \texttt{EAZY} \citep{EAZY}, a widely used code for estimating photometric redshifts by fitting observed data with template libraries.
In this study, we apply the following three fitting setups using public templates:
\begin{itemize}
    \item Fitting with a \texttt{sorted\_agn\_blue\_sfhz\_13} template set, a template set for LRDs constructed based on the spectrum in \cite{Killi2024} and available in \texttt{EAZY}, using a redshift range of $z = 0\,\mathchar`-\,15$ with a step of $\Delta z=0.01$.
    \item Fitting with a galaxy template set, \texttt{tweak\_fsps\_QSF\_12\_v3}, using a redshift range of $z = 2\,\mathchar`-\,15$ with a step of $\Delta z=0.01$ for specifically testing the high-$z$ non-LRD galaxy case.
    \item Fitting with a galaxy template set, \texttt{tweak\_fsps\_QSF\_12\_v3}, using a redshift range of $z = 0\,\mathchar`-\,2$ with a step of $\Delta z=0.01$ for specifically testing the $z \sim 0.2$ interloper case.
\end{itemize}
One huge difference between \texttt{EAZY} and \texttt{BAGPIPES} (Section\,\ref{ss:SED_fit}) is that \texttt{BAGPIPES} assumes a single galaxy population, in contrast to \texttt{EAZY}, which fits composite SEDs from multiple templates.
The $\chi^2$ distribution and the best-fit model SED are shown in Figure\,\ref{fig:sedfit_eazy}.

When fitting over the wide redshift range of $0<z<15$ with the \texttt{sorted\_agn\_blue\_sfhz\_13} (LRD) template set, we obtain a best-fit photometric redshift of $z_{\rm photo}=10.13^{+0.60}_{-0.26}$.
Then, when fitting over the low redshift range of $0<z<2$ with \texttt{tweak\_fsps\_QSF\_12\_v3} (galaxy) template, we obtain a best-fit photometric redshift of $z_{\rm photo}=0.22^{+0.02}_{-0.01}$.
Thus, the $z\sim10$ LRD solution has much lower $\chi^2$ with $\Delta \chi^2=15.6$ than the $z\sim0.2$ galaxy case, consistent with the results with \texttt{BAGPIPES} (Section\,\ref{ss:photo-z}).
Also note that the inferred photometric redshift with the LRD template ($z_{\rm photo} = 10.13^{+0.60}_{-0.26}$) is consistent with the result obtained using the \citet{Akins2024} model ($z_{\rm photo} = 10.50^{+0.72}_{-0.58}$) within the 1$\sigma$ confidence level.

Fitting over the high-redshift range of $2 < z < 15$ with the \texttt{tweak\_fsps\_QSF\_12\_v3} (galaxy) template yields a photometric redshift of $z_{\rm photo} = 10.82^{+0.31}_{-0.55}$.
The $z \sim 10$ galaxy fitting has a higher $\chi^2$ value, with $\Delta \chi^2 = 7.07$, compared to the $z \sim 10$ LRD solution.
A major difference from the \texttt{BAGPIPES} results is that, with \texttt{EAZY}, the $z \sim 10$ galaxy case can simultaneously reproduce both the rest-frame UV and rest-frame optical parts of the SED, resulting in a lower $\chi^2$ than the $z\sim10$ galaxy model fit obtained with \texttt{BAGPIPES} (Figure\,\ref{fig:sedfit} green line).
This improved fitting result arises because \texttt{EAZY} models SEDs as a linear combination of multiple templates, each potentially characterized by different physical parameters.
In this case, the best-fit SED consists of a combination of a dust-unobscured SFG ($A_V=0.068$, $M_*=2.5 \times 10^{10}M_\odot$, and ${\rm SFR}=4.8M_\odot\,\mathrm{yr}^{-1}$) and a dust-obscured massive quiescent galaxy ($A_V=2.884$, $M_*=6.2 \times 10^{11}M_\odot$, and ${\rm SFR}\simeq0M_\odot\,\mathrm{yr}^{-1}$), each contributing primarily to the rest-frame UV and rest-frame optical parts of the SED, respectively.
Notably, the total inferred stellar mass ($M_* = 6.5\times 10^{11}M_\odot$) is too large for a galaxy at $z\sim10$ and is not allowed in the framework of the $\Lambda$CDM, even assuming a global star formation efficiency of $\epsilon=1$ \citep{Boylan_Kolchin2023}.
Furthermore, when combined with the upper limit on the galaxy size ($< 120\,{\rm pc}$, Section\,\ref{ss:morph}), the inferred $M_*$ implies a lower limit on the stellar mass density of $\Sigma_* >7.2\times10^{12}M_\odot\,{\rm kpc^{-2}}$.
This value far exceeds the maximum surface densities observed in compact stellar systems in the local Universe ($\Sigma_* \sim3\times10^{11}M_\odot\,{\rm kpc^{-2}}$, \citealt{Hopkins2010, Grudic2019}), making it physically implausible.

Thus, even based on the fitting results from \texttt{EAZY}, we conclude that the most plausible interpretation of CW-LRD-z10 is an LRD at $z_{\rm photo}=10.13^{+0.60}_{-0.26}$.
With this photometric redshift, the inferred UV absolute magnitude is $M_{\rm UV} = -19.9\pm0.1$, consistent with the result derived from the \texttt{BAGPIPES} fitting (Section\,\ref{ss:photo-z}).

\bibliography{export}{}
\bibliographystyle{aasjournalv7}

\allauthors


\end{document}

%% file: author.tex
\author[0009-0003-4742-7060]{Takumi S. Tanaka}
\affiliation{Kavli Institute for the Physics and Mathematics of the Universe (WPI), The University of Tokyo Institutes for Advanced Study, The University of Tokyo, Kashiwa, Chiba 277-8583, Japan}
\affiliation{Department of Astronomy, Graduate School of Science, The University of Tokyo, 7-3-1 Hongo, Bunkyo-ku, Tokyo 113-0033, Japan}
\affiliation{Center for Data-Driven Discovery, Kavli IPMU (WPI), UTIAS, The University of Tokyo, Kashiwa, Chiba 277-8583, Japan}
\email[show]{takumi.tanaka@ipmu.jp}
\author[0000-0003-3596-8794]{Hollis B. Akins}
\affiliation{The University of Texas at Austin, 2515 Speedway Blvd Stop C1400, Austin, TX 78712, USA}
\email{hollis.akins@gmail.com}
\author[0000-0002-6047-430X]{Yuichi Harikane}
\affiliation{Institute for Cosmic Ray Research, The University of Tokyo, 5-1-5 Kashiwanoha, Kashiwa, Chiba 277-8582, Japan}
\email{hari@icrr.u-tokyo.ac.jp}
\author[0000-0002-0000-6977]{John D. Silverman}
\affiliation{Kavli Institute for the Physics and Mathematics of the Universe (WPI), The University of Tokyo Institutes for Advanced Study, The University of Tokyo, Kashiwa, Chiba 277-8583, Japan}
\affiliation{Department of Astronomy, Graduate School of Science, The University of Tokyo, 7-3-1 Hongo, Bunkyo-ku, Tokyo 113-0033, Japan}
\affiliation{Center for Data-Driven Discovery, Kavli IPMU (WPI), UTIAS, The University of Tokyo, Kashiwa, Chiba 277-8583, Japan}
\email{john.silverman@ipmu.jp}
\author[0000-0002-0930-6466]{Caitlin M. Casey}
\affiliation{Department of Physics, University of California, Santa Barbara, Santa Barbara, CA 93106, USA}
\affiliation{The University of Texas at Austin, 2515 Speedway Blvd Stop C1400, Austin, TX 78712, USA}
\affiliation{Cosmic Dawn Center (DAWN), Denmark}
\email{cmcasey@ucsb.edu}
\author[0000-0001-9840-4959]{Kohei Inayoshi}
\affiliation{Kavli Institute for Astronomy and Astrophysics, Peking University, Beijing 100871, China}
\email{inayoshi.pku@gmail.com}
\author[orcid=0000-0002-4544-8242]{Jan-Torge Schindler}
\affiliation{Hamburger Sternwarte, University of Hamburg, Gojenbergsweg 112, D-21029 Hamburg, Germany}
\email{jan-torge.schindler@uni-hamburg.de}
\author[0000-0002-2597-2231]{Kazuhiro Shimasaku}
\affiliation{Department of Astronomy, Graduate School of Science, The University of Tokyo, 7-3-1 Hongo, Bunkyo-ku, Tokyo 113-0033, Japan}
\affiliation{Research Center for the Early Universe, Graduate School of Science, The University of Tokyo, 7-3-1 Hongo, Bunkyo-ku, Tokyo 113-0033, Japan}
\email{shimasaku@astron.s.u-tokyo.ac.jp}
%
%
\author[0000-0002-8360-3880]{Dale D. Kocevski}
\affiliation{Department of Physics and Astronomy, Colby College, Waterville, ME 04901, USA}
\email{dkocevsk@colby.edu}
\author[0000-0003-2984-6803]{Masafusa Onoue}
\affiliation{Waseda Institute for Advanced Study (WIAS), Waseda University, 1-21-1, Nishi-Waseda, Shinjuku, Tokyo 169-0051, Japan; Center for Data Science, Waseda University, 1-6-1, Nishi-Waseda, Shinjuku, Tokyo 169-0051, Japan}
\affiliation{Kavli Institute for the Physics and Mathematics of the Universe (WPI), The University of Tokyo Institutes for Advanced Study, The University of Tokyo, Kashiwa, Chiba 277-8583, Japan}
\email{masafusa.onoue@aoni.waseda.jp}
\author[0000-0002-9382-9832]{Andreas L. Faisst}
\affiliation{Caltech/IPAC, MS 314-6, 1200 E. California Blvd. Pasadena, CA 91125, USA}
\email{afaisst@ipac.caltech.edu}
\author[0000-0002-4271-0364]{Brant E. Robertson}
\affiliation{Department of Astronomy and Astrophysics, University of California, Santa Cruz, 1156 High Street, Santa Cruz, CA 95064, USA}
\email{brant@ucsc.edu}
\author[0000-0002-5588-9156]{Vasily Kokorev}
\affiliation{Kapteyn Astronomical Institute, University of Groningen, PO Box 800, 9700 AV Groningen, The Netherlands}
\email{vasily.kokorev.astro@gmail.com}
\author[0000-0002-7087-0701]{Marko Shuntov}
\affiliation{Cosmic Dawn Center (DAWN), Denmark} 
\affiliation{Niels Bohr Institute, University of Copenhagen, Jagtvej 128, DK-2200, Copenhagen, Denmark}
\affiliation{University of Geneva, 24 rue du Général-Dufour, 1211 Genève 4, Switzerland}
\email{marko.shuntov@nbi.ku.dk}
\author[0000-0002-6610-2048]{Anton M. Koekemoer}
\affiliation{Space Telescope Science Institute, 3700 San Martin Dr., Baltimore, MD 21218, USA} 
\email{koekemoer@stsci.edu}
\author[0000-0002-3560-8599]{Maximilien Franco}
\affiliation{Université Paris-Saclay, Université Paris Cité, CEA, CNRS, AIM, 91191 Gif-sur-Yvette, France}
\email{maximilien.franco@cea.fr}
\author[0000-0003-1344-9475]{Eiichi Egami}
\affiliation{
Steward Observatory, University of Arizona, 933 N. Cherry Ave, Tucson, AZ 85721, USA
}
\email{egami@arizona.edu}
\author[0000-0001-9773-7479]{Daizhong Liu}
\affiliation{Purple Mountain Observatory, Chinese Academy of Sciences, 10 Yuanhua Road, Nanjing 210023, China}
\email{dzliu@pmo.ac.cn}
\author[0000-0003-1282-7454]{Anthony J. Taylor}
\affiliation{Department of Astronomy, The University of Texas at Austin, 2515 Speedway, Austin, TX, 78712, USA}
\email{anthony.taylor@austin.utexas.edu}
\author[0000-0001-9187-3605]{Jeyhan S. Kartaltepe}
\affiliation{Laboratory for Multiwavelength Astrophysics, School of Physics and Astronomy, Rochester Institute of Technology, 84 Lomb Memorial Drive, Rochester, NY 14623, USA}
\email{jeyhan@astro.rit.edu}
\author[0000-0001-8582-7012]{Sarah E.~I.~Bosman}
\affiliation{Institute for Theoretical Physics, Heidelberg University, Philosophenweg 12, D–69120, Heidelberg, Germany}
\affiliation{Max-Planck-Institut f\"{u}r Astronomie, K\"{o}nigstuhl 17, 69117 Heidelberg, Germany}
\email{bosman@thphys.uni-heidelberg.de}
\author[0000-0002-6184-9097]{Jaclyn B. Champagne}
\affiliation{Steward Observatory, University of Arizona, 933 N Cherry Avenue, Tucson, AZ 85721, USA}
\email{jbchampagne@arizona.edu}
\author[0000-0001-6874-1321]{Koki Kakiichi}
\affiliation{Cosmic Dawn Center (DAWN), Denmark}
\affiliation{Niels Bohr Institute, University of Copenhagen, Jagtvej 128, DK-2200, Copenhagen, Denmark}
\email{koki.kakiichi@nbi.ku.dk}
\author[0000-0003-0129-2079]{Santosh Harish}
\affiliation{Laboratory for Multiwavelength Astrophysics, School of Physics and Astronomy, Rochester Institute of Technology, 84 Lomb Memorial Drive, Rochester, NY 14623, USA}
\email{harish.santosh@gmail.com}
%
%
\author[0000-0002-2420-5022]{Zijian Zhang}
\affiliation{Kavli Institute for Astronomy and Astrophysics, Peking University, Beijing 100871, China}
\affiliation{Department of Astronomy, School of Physics, Peking University, Beijing 100871, China}
\email{zzhang.astro@gmail.com}
\author[0009-0001-3422-3048]{Sophie L. Newman}
\affiliation{Institute of Cosmology and Gravitation, University of Portsmouth, Dennis Sciama Building, Burnaby Road, Portsmouth, PO1 3FX, UK}
\email{sophie.newman@port.ac.uk}
\author[0000-0002-2603-2639]{Darshan Kakkad}
\affiliation{Centre for Astrophysics Research, University of Hertfordshire, Hatfield, AL10 9AB, UK}
\email{darshankakkad@gmail.com}
\author[0000-0001-7232-5355]{Qinyue Fei}
\affiliation{David A. Dunlap Department of Astronomy and Astrophysics, University of Toronto, 50 St. George Street, Toronto, Ontario, M5S 3H4, Canada}
\email{qyfei.astro@gmail.com}
\author[0000-0001-7201-5066]{Seiji Fujimoto}
\affiliation{Department of Astronomy, The University of Texas at Austin, 2515 Speedway, Austin, TX, 78712, USA}
\affiliation{David A. Dunlap Department of Astronomy and Astrophysics, University of Toronto, 50 St. George Street, Toronto, Ontario, M5S 3H4, Canada}
\affiliation{Dunlap Institute for Astronomy and Astrophysics, 50 St. George Street, Toronto, Ontario, M5S 3H4, Canada}
\email{seiji.fujimoto@astro.utoronto.ca}
\author[0000-0001-6251-649X]{Mingyu Li}
\affiliation{Department of Astronomy, Tsinghua University, Beijing 100084, China}
\email{lmytime@hotmail.com}
\author[0000-0001-8519-1130]{Steven L. Finkelstein}
\affiliation{Department of Astronomy, The University of Texas at Austin, 2515 Speedway, Austin, TX, 78712, USA}
\email{stevenf@astro.as.utexas.edu}
\author[0000-0001-7634-1547]{Zi-Jian Li}
\affiliation{Chinese Academy of Sciences South America Center for Astronomy (CASSACA), \\
National Astronomical Observatories of China (NAOC),
CAS, 20A Datun Road, Beijing 100012, China}
\affiliation{School of Astronomy and Space Sciences, University of Chinese Academy of Sciences, Beijing 100049, China}
\email{zjli@nao.cas.cn}
\author[0000-0003-3216-7190]{Erini Lambrides}
\affiliation{NASA-Goddard Space Flight Center, Code 662, Greenbelt, MD, 20771, USA}
\email{elambrid@gmail.com}
\author[0000-0002-2906-2200]{Laura Sommovigo}
\affiliation{Center for Computational Astrophysics, Flatiron Institute, 162 5th Avenue, New York, NY 10010, USA}
\email{laura.sommovigo.work@gmail.com}
\author[0000-0002-7051-1100]{Jorge A. Zavala}
\affiliation{University of Massachusetts Amherst, 710 North Pleasant Street, Amherst, MA 01003-9305, USA}
\email{jorge.zavala@nao.ac.jp}
\author[0000-0002-9453-0381]{Kei Ito}
\affiliation{Cosmic Dawn Center (DAWN), Denmark}
\affiliation{DTU Space, Technical University of Denmark, Elektrovej 327 DK2800 Kgs. Lyngby, Denmark}
\email{keiitoastro@gmail.com}
\author[0000-0002-9252-114X]{Zhaoxuan Liu}
\affiliation{Kavli Institute for the Physics and Mathematics of the Universe (WPI), The University of Tokyo Institutes for Advanced Study, The University of Tokyo, Kashiwa, Chiba 277-8583, Japan}
\affiliation{Department of Astronomy, Graduate School of Science, The University of Tokyo, 7-3-1 Hongo, Bunkyo-ku, Tokyo 113-0033, Japan}
\affiliation{Center for Data-Driven Discovery, Kavli IPMU (WPI), UTIAS, The University of Tokyo, Kashiwa, Chiba 277-8583, Japan}
\email{zhaoxuan.liu@ipmu.jp}
\author[0000-0001-7568-6412]{Ezequiel Treister}
\affiliation{Instituto de Alta Investigaci{\'{o}}n, Universidad de Tarapac{\'{a}}, Casilla 7D, Arica, Chile}
\email{etreister@academicos.uta.cl}
\author[0000-0002-6290-3198]{Manuel Aravena}
\affiliation{Instituto de Estudios Astrof\'{\i}cos, Facultad de Ingenier\'{\i}a y Ciencias, Universidad Diego Portales, Av. Ej\'ercito 441, Santiago, Chile}
\affiliation{Millenium Nucleus for Galaxies (MINGAL)}
\email{manuel.aravenaa@mail.udp.cl}
\author[0000-0002-0236-919X]{Ghassem Gozaliasl}
\affiliation{Department of Computer Science, Aalto University, P.O. Box 15400, FI-00076 Espoo, Finland}
\affiliation{Department of Physics, University of, P.O. Box 64, FI-00014 Helsinki, Finland}
\email{ghassemgozaliasl@gmail.com}
\author[0000-0002-4321-3538]{Haowen Zhang}
\affiliation{Steward Observatory, University of Arizona, 933 N Cherry Avenue, Tucson, AZ 85721, USA}
\email{hwzhang0595@arizona.edu}
\author[0009-0007-3673-4523]{Hossein Hatamnia}
\affiliation{Department of Physics and Astronomy, University of California, Riverside, 900 University Avenue, Riverside, CA 92521, USA}
\email{hhata003@ucr.edu}
\author[0009-0008-0167-5129]{Hiroya Umeda}
\affiliation{Institute for Cosmic Ray Research, The University of Tokyo, 5-1-5 Kashiwanoha, Kashiwa, Chiba 277-8582, Japan}
\affiliation{Department of Physics, Graduate School of Science, The University of Tokyo, 7-3-1 Hongo, Bunkyo, Tokyo 113-0033, Japan}
\email{ume@icrr.u-tokyo.ac.jp}
\author[0000-0002-7779-8677]{Akio K. Inoue}
\affiliation{Department of Pure and Applied Physics, Graduate School of Advanced Science and Engineering, Faculty of Science and Engineering, Waseda University, 3-4-1, Okubo, Shinjuku, Tokyo, 169-8555, Japan}
\affiliation{Waseda Research Institute for Science and Engineering, Faculty of Science and Engineering, Waseda University, 3-4-1, Okubo, Shinjuku, Tokyo, 169-8555, Japan}
\email{akinoue@aoni.waseda.jp}
\author[0000-0001-5287-4242]{Jinyi Yang}
\affiliation{Department of Astronomy, University of Michigan, 1085 S. University Ave., Ann Arbor, MI 48109, USA}
\email{jyyangas@umich.edu}
\author[0000-0002-4225-4477]{Makoto Ando}
\affiliation{National Astronomical Observatory of Japan, 2-21-1 Osawa, Mitaka, Tokyo, 181-8588, Japan}
\email{makoto.ando.astro@gmail.com}
\author[0009-0007-0864-7094]{Junya Arita}
\affiliation{Department of Astronomy, Graduate School of Science, The University of Tokyo, 7-3-1 Hongo, Bunkyo-ku, Tokyo 113-0033, Japan}
\email{jarita@astron.s.u-tokyo.ac.jp}
\author[0000-0001-8917-2148]{Xuheng Ding}
\affiliation{School of Physics and Technology, Wuhan University, Wuhan 430072, China}
\email{dingxh@whu.edu.cn}
\author[orcid=0009-0009-7403-8603]{Suin Matsui}
\affiliation{Department of Astronomy, Graduate School of Science, The University of Tokyo, 7-3-1 Hongo, Bunkyo-ku, Tokyo 113-0033, Japan}
\email{smatsui@astron.s.u-tokyo.ac.jp}
\author[orcid=0009-0005-7449-8184]{Yuki Shibanuma}
\affiliation{Department of Astronomy, Graduate School of Science, The University of Tokyo, 7-3-1 Hongo, Bunkyo-ku, Tokyo 113-0033, Japan}
\email{shibanuma@astron.s.u-tokyo.ac.jp}
\author[0000-0002-4872-2294]{Georgios Magdis}
\affiliation{Cosmic Dawn Center (DAWN), Denmark} 
\affiliation{DTU Space, Technical University of Denmark, Elektrovej 327 DK2800 Kgs. Lyngby, Denmark}
\affiliation{Niels Bohr Institute, University of Copenhagen, Jagtvej 128, DK-2200, Copenhagen, Denmark}
\email{georgios.magdis@gmail.com}
%
%
\author[0000-0001-5105-2837]{Mingyang Zhuang}
\affiliation{Department of Astronomy, University of Illinois at Urbana-Champaign, Urbana, IL 61801, USA}
\email{mingyang@illinois.edu}
\author[0000-0003-3310-0131]{Xiaohui Fan}
\affiliation{Steward Observatory, University of Arizona, 933 N Cherry Avenue, Tucson, AZ 85721, USA}
\email{xfan@arizona.edu}
\author[0000-0001-5951-459X]{Zihao Li}
\affiliation{Cosmic Dawn Center (DAWN), Denmark}
\affiliation{Niels Bohr Institute, University of Copenhagen, Jagtvej 128, DK-2200, Copenhagen N, Denmark}
\email{zihao.li@nbi.ku.dk}
\author[0000-0003-3762-7344]{Weizhe Liu} 
\affiliation{Steward Observatory, University of Arizona, 933 N Cherry Avenue, Tucson, AZ 85721, USA}
\email{wzliu@arizona.edu}
\author[0000-0002-6221-1829]{Jianwei Lyu}
\affiliation{Steward Observatory, University of Arizona, 933 N Cherry Avenue, Tucson, AZ 85721, USA}
\email{jianwei@arizona.edu}
\author[0000-0002-4485-8549]{Jason Rhodes}
\affiliation{Jet Propulsion Laboratory, California Institute of Technology, 4800 Oak Grove Drive, Pasadena, CA 91001, USA}
\email{jason.d.rhodes@jpl.nasa.gov}
\author[0000-0003-3631-7176]{Sune Toft}
\affiliation{Cosmic Dawn Center (DAWN), Denmark} 
\affiliation{Niels Bohr Institute, University of Copenhagen, Jagtvej 128, DK-2200, Copenhagen, Denmark}
\email{sune@nbi.ku.dk}
\author[0000-0002-7633-431X]{Feige Wang}
\affiliation{Department of Astronomy, University of Michigan, 1085 S. University Ave., Ann Arbor, MI 48109, USA}
\email{fgwang@umich.edu}
\author[0000-0002-3983-6484]{Siwei Zou}
\affiliation{Chinese Academy of Sciences South America Center for Astronomy, National Astronomical Observatories, CAS, Beijing 100101, China}
\affiliation{Departamento de Astronom\'ia, Universidad de Chile, Casilla 36-D, Santiago, Chile}
\email{siwei1905@gmail.com}
%

%
\author[0000-0002-0569-5222]{Rafael C. Arango-Toro}
\affiliation{Aix Marseille Univ, CNRS, CNES, LAM, Marseille, France}
\email{rafael.arango-toro@lam.fr}
\author[0000-0003-4569-2285]{Andrew J. Battisti}
\affiliation{Research School of Astronomy and Astrophysics, Australian National University, Cotter Road, Weston Creek, ACT 2611, Australia}
\affiliation{ARC Centre of Excellence for All Sky Astrophysics in 3 Dimensions (ASTRO 3D), Australia}
\email{Andrew.Battisti@anu.edu.au}
\author[0000-0001-9885-4589]{Steven Gillman}
\affiliation{Cosmic Dawn Center (DAWN), Denmark}
\affiliation{DTU Space, Technical University of Denmark, Elektrovej 327 DK2800 Kgs. Lyngby, Denmark}
\email{srigi@space.dtu.dk}
\author[0000-0002-0101-336X]{Ali Ahmad Khostovan}
\affiliation{Department of Physics and Astronomy, University of Kentucky, 505 Rose Street, Lexington, KY 40506, USA}
\affiliation{Laboratory for Multiwavelength Astrophysics, School of Physics and Astronomy, Rochester Institute of Technology, 84 Lomb Memorial Drive, Rochester, NY 14623, USA}
\email{akhostov@gmail.com}
\author[0000-0002-7530-8857]{Arianna S. Long}
\affiliation{Department of Astronomy, The University of Washington, Seattle, WA 98195, USA}
\email{arianna.sage.long@gmail.com}
\author[0000-0001-5846-4404]{Bahram Mobasher}
\affiliation{Department of Physics and Astronomy, University of CaliforniaPhysics and Astronomy Department, University of California, 900 University Ave., Riverside, CA 92521, USA}
\email{mobasher@ucr.edu}
\author[0000-0002-1233-9998]{David B. Sanders}
\affiliation{Institute for Astronomy, University of Hawai’i at Manoa, 2680 Woodlawn Drive, Honolulu, HI 96822, USA}
\email{sanders@ifa.hawaii.edu} 